\documentclass{ws-procs11x85}
\usepackage{balance}
\usepackage{hyperref}
\usepackage{graphicx}

\makeindex
\begin{document}

\title{OUTLOOK: THE NEXT TWENTY YEARS}

\author{H.~MURAYAMA}

\address{Department of Physics, University of California, Berkeley, CA
94720\\
and\\
Theoretical Physics Groups, Lawrence Berkeley Laboratory,
University of California, Berkeley, CA 94720\\
and\\
School of Natural Sciences, Institute for Advanced Study,
Princeton, NJ 08540 \\E-mail: murayama@ias.edu}


\twocolumn[\maketitle\abstract{I present an outlook for the next
  twenty years in particle physics.  I start with the big questions in
  our field, broken down into four categories: horizontal, vertical,
  heaven, and hell.  Then I discuss how we attack the big questions in
  each category during the next twenty years.  I argue for a synergy
  between many different approaches taken in our field.}]

\baselineskip=13.07pt

\section{Introduction}

I was asked specifically not to give a summary talk of the Symposium,
but rather an ``outlook'' talk.  Obviously I will refer to many
excellent talks given during this Symposium, but I will not try to
cover everything that had been said.  If you find that your favorite
subject is not covered in my talk, you are welcome to come up here and
complain to Keith Ellis.

\begin{figure}[t]
  \center
  \includegraphics[width=\columnwidth]{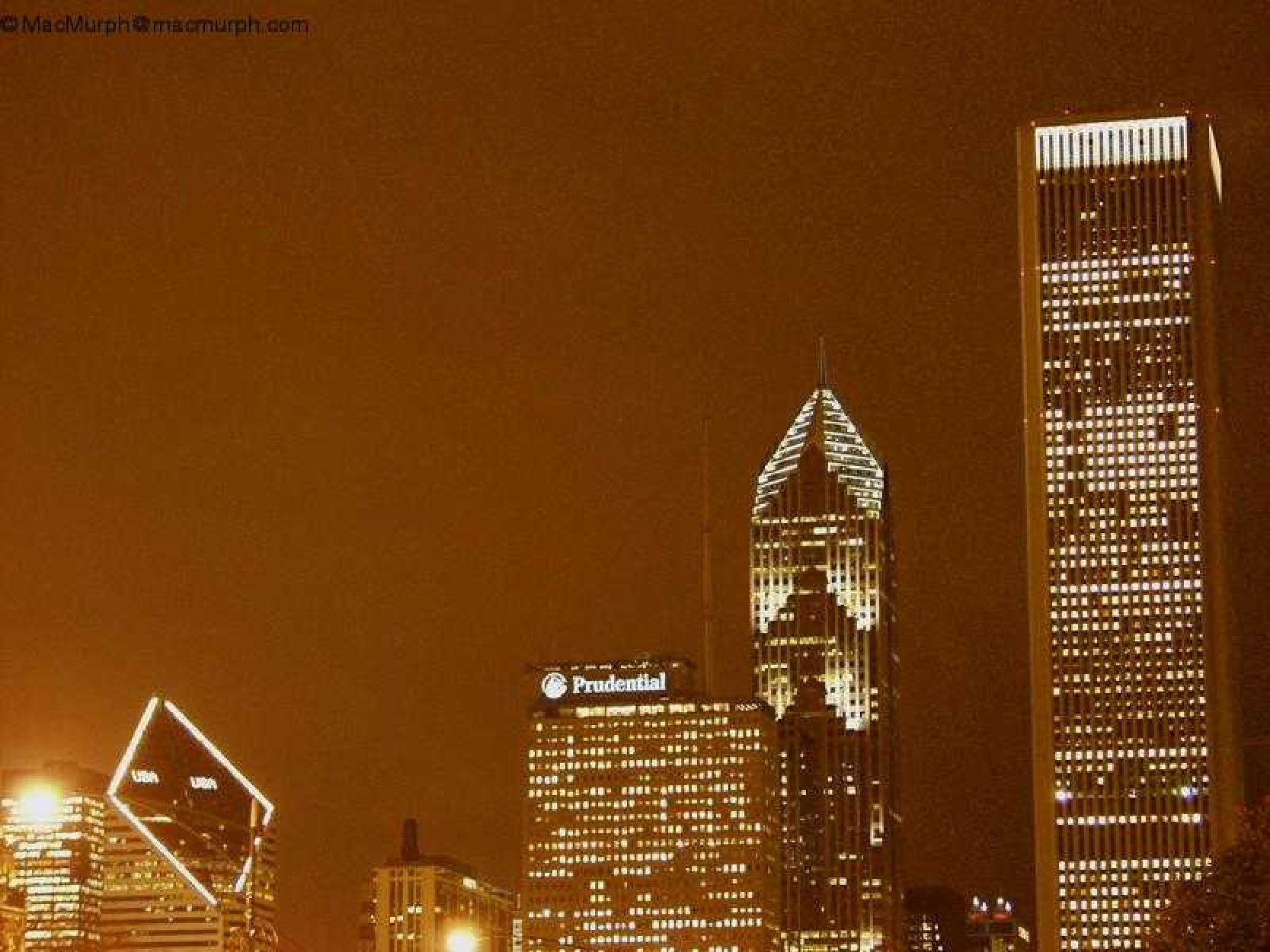}
  \caption{The bright skyline of Chicago.\label{fig:Chicago}}
\vspace{-0.4cm}
\end{figure}

\begin{figure}[h!]
  \center
  \includegraphics[width=\columnwidth]{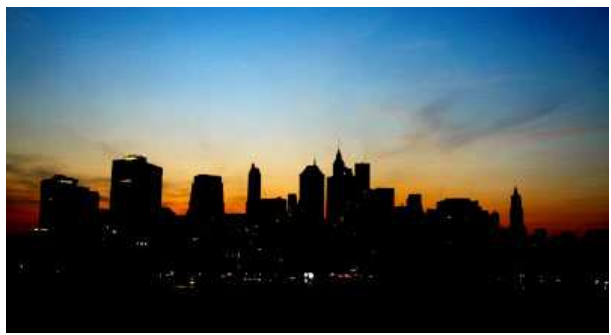}
  \caption{The dark skyline of New York City during the
    blackout.\label{fig:NewYork}}
\vspace{-0.4cm}
\end{figure}

What I will try to do in this talk is present my own perspective on
how our field may develop in the next twenty years.  Whenever we talk
about the future, a very natural question is whether it is bright, as
bright as the illumination of the Chicago skyscrapers we admired at the
time of the banquet (Fig.~\ref{fig:Chicago}), or dark, as dark as
parts of the East Coast were this week (Fig.~\ref{fig:NewYork}).  You
will see my verdict at the end of the talk.  But before getting to the
verdict, I'd like to talk about the current situation of our field.


One way to look at our field of particle physics is that it has
specialized into so many different subfields.  During this Symposium
alone, we heard about many exciting subjects and experiments.  Here is
an {\it incomplete}\/ list in a random order: $0\nu\beta\beta$,
$B$-physics, proton decay, LHC, Higgs, reactor anti-neutrinos,
$K$-physics, Lepton Flavor Violation, Cosmic Microwave Background,
string theory, $e^+ e^-$, top quark, accelerator-based neutrinos,
lattice QCD, charm physics, Dark Energy, hidden dimensions, neutrino
factory, hadron physics, Supersymmetry, Dark Matter, atmospheric
neutrinos, Linear Collider, exotics, solar neutrinos, and the increasingly
important topics of outreach and politics.  People may say that the
field is completely fragmented.  On the other hand, I have a somewhat
different view.  I think that most of them are heading to a synergy at
an important energy scale: TeV.  I don't mean that TeV-scale physics
will solve all the puzzles we are facing.  What I mean is that any of
these interesting physics topics must once go through the study of the
TeV scale before they reach their own destinations.  It is a hub where
everybody has to transfer to another flight.

Why do I think so?  To see this, let me start enumerating the big
questions in our field.  I broke them down to four categories.

\begin{figure}[t]
  \center
  \includegraphics[width=0.7\columnwidth]{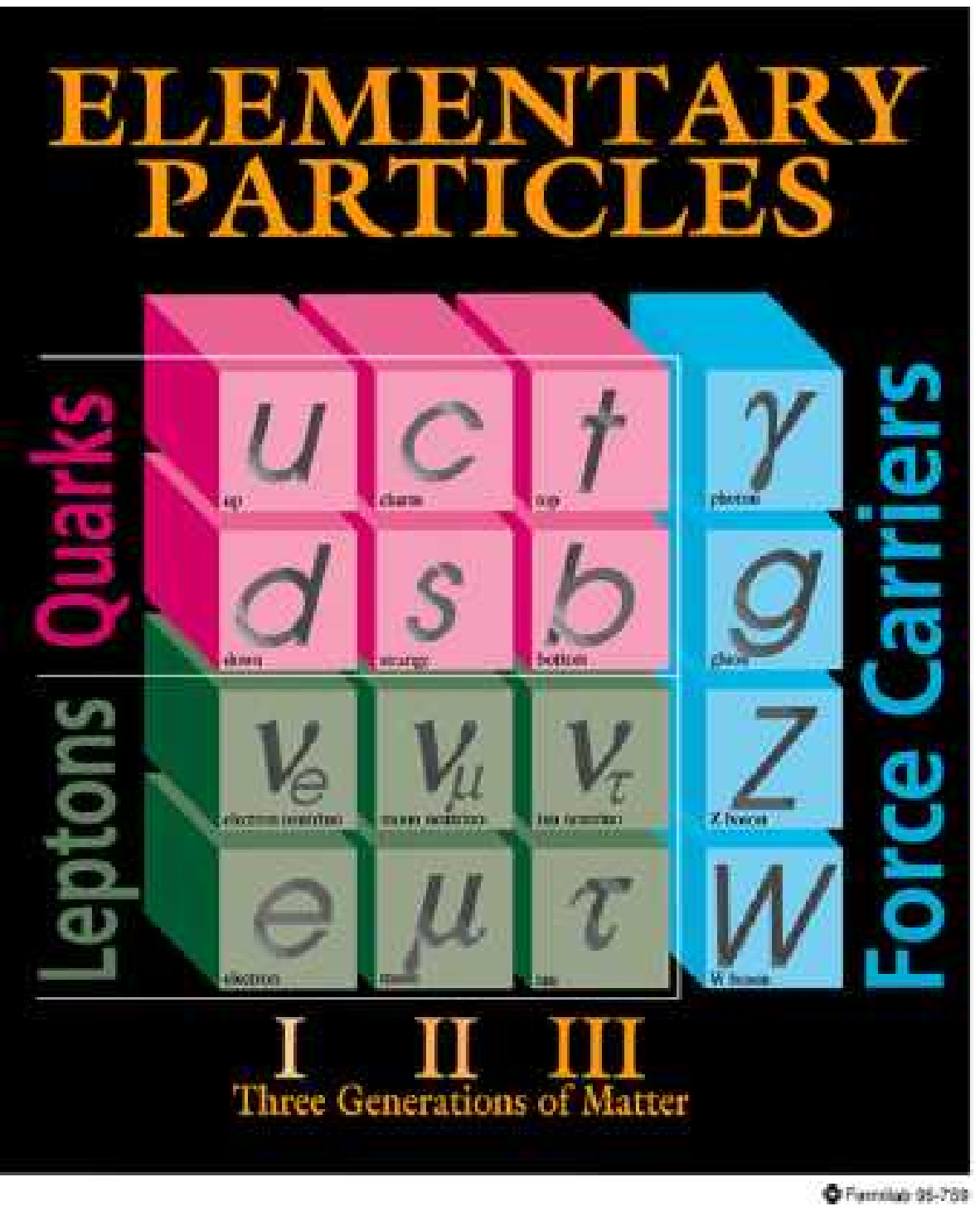}
  \caption[]{The table of particles in the Standard
    Model\rlap.\,\cite{SM} Vertical Questions are concerned with particles
    in a single generation, while Horizontal Questions refer to the
    relationship sideways in the table.\label{fig:chart}}
\end{figure}

\begin{figure}
  \center
  \includegraphics[width=\columnwidth]{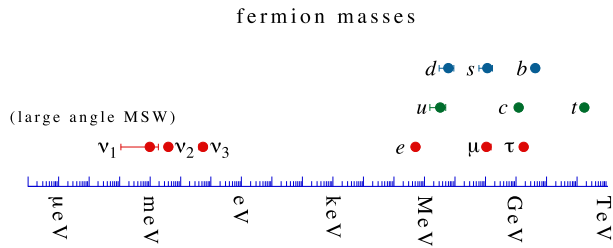}
  \caption{The mass spectrum of quarks and leptons we don't
    understand. \label{fig:masses}} 
\end{figure}

The first category is what I call the {\it horizontal questions}\/.
They are about relationships between the three families of elementary
particles (Fig.~\ref{fig:chart}).  Why are there three generations and
no more?  What physics determines their masses and mixings
(Fig.~\ref{fig:masses})? What is the energy scale of that physics?  Why
do neutrinos have mass and yet they are so light?  What is the origin of
CP-violation?  What is the origin of the matter-antimatter asymmetry
in our Universe?


The second category is the {\it vertical questions}\/.  They concern
properties within each family of particles (Fig.~\ref{fig:chart}).
Why are there three unrelated gauge forces?  Why is the strong
interaction strong? Why are all electric charges quantized in the same
unit?  What physics guarantees the seemingly miraculous anomaly
cancellation?  What physics explains the quantum numbers of quarks and
leptons we see (Table~\ref{tab:quantumnumbers})? Is there a unified
description of all forces? Why is $m_W \ll M_{\rm Planck}$? (Hierarchy
Problem.)

\begin{table}
\center
\begin{eqnarray*}
  & &
  Q({\bf 3}, {\bf 2}, +\frac{1}{6}), \qquad
  u(\bar{\bf 3}, {\bf 1}, -\frac{2}{3}), \qquad
  d(\bar{\bf 3}, {\bf 1}, +\frac{1}{3}), \\
  & &
  L({\bf 1}, {\bf 2}, -\frac{1}{2}), \qquad
  e({\bf 1}, {\bf 1}, +1)
\end{eqnarray*}
\caption{The quantum numbers of quarks and leptons we don't
  understand.  Here, all particles are shown in their left-handed
  chirality states.\label{tab:quantumnumbers}}
\end{table}

\begin{figure}[t]
  \center
  \includegraphics[width=\columnwidth]{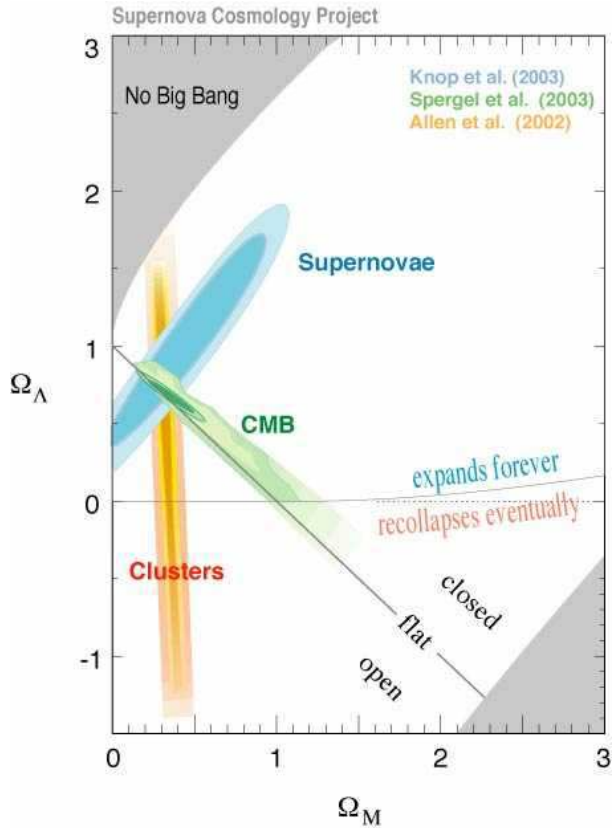}
  \caption{The unknown constituents of the universe.\label{fig:makeup}}
\vspace{-0.5cm}
\end{figure}

Recently, we added many {\it questions from the heaven}
(Fig.~\ref{fig:makeup}). What is Dark Matter?  What is Dark Energy?
Why are we at the special moment when the energy densities of Dark
Matter and Dark Energy are the same within a factor of two?  (``Why
now?'' problem.) What exactly was the Big Bang?  Why is the Universe so
big? (Flatness problem, horizon problem.) How were galaxies and stars
(and eventually us) created?

\begin{figure}[t]
  \center
  \includegraphics[width=0.7\columnwidth]{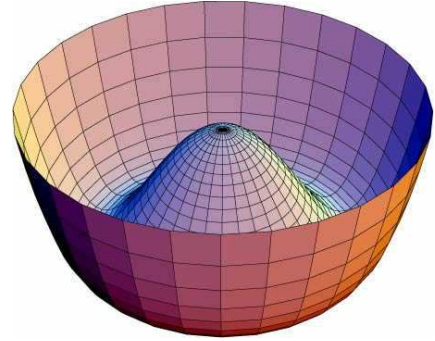}
  \caption{The hell of the Universe we live in and don't know
    why.\label{fig:hell}} 
\vspace{-0.5cm}
\end{figure}

If there are questions from the heaven, there are also {\it questions
  from the hell}\/.  To the best of the collective knowledge of Homo
sapiens, we live at the bottom of a strange potential with a
wine bottle shape: that's the hell we are in (Fig.~\ref{fig:hell}).
Because of this potential, the Bose--Einstein condensate (BEC) of the Higgs
boson is supposed to be present in our Universe, and we are swimming in
this BEC.  What is this Higgs boson thing?  Why does it have this
strange potential with a negative mass-squared?  Why is there only one
scalar particle in the Standard Model, designed to do its most
mysterious part?  Is it elementary or composite?  Is it {\it really}\/
condensed in our Universe?

We do not have the right to expect that {\it any}\/ of these questions
can be answered within our lifetime (or ever).  Nonetheless there is a
good potential for us to answer some or many of them.  How exactly do
we do it?  I will refer to Supersymmetry as an example many times in
my talk, but I expect similar stories with any scenario of TeV-scale
physics.  In any case, TeV is the key.

I have just finished the introduction.  Now I move on to discuss each
category of questions: hell, heaven, vertical and horizontal.  My
verdict on the future of our field follows after that.

\section{Hell}

\begin{figure}
\center
\includegraphics[width=0.5\columnwidth]{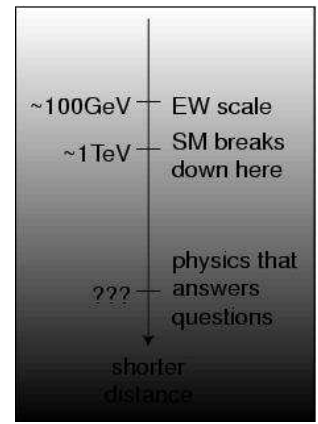}
\caption{We would like to access physics at a short distance that
  answers some of the big questions.  But before getting there, the
  Standard Model breaks down around a TeV and everything at shorter
  distances is grayed out.\label{fig:shade}}
\end{figure}

What we know is that the Standard Model of particle physics is
completely incapable of answering the big questions I've listed.  What
we want to do is to look for physics beyond the Standard Model that
answers these big questions.  By definition, that is physics at
shorter distances.  In order to talk about the new physics that
appears at a some small distance scale, the Standard Model must
survive down to whatever that short distance scale is.  The problem is
that it doesn't.  This is the hierarchy problem.  It is the main
obstacle for us to address the big questions. We can't even get
started! (Fig.~\ref{fig:shade})

To illustrate the reason why we can't even get started, let us
rewind the video back to the end of the 19th century.  {\it Once upon
  a time}\/, there was a hierarchy problem\rlap.\,\cite{Murayama:1994kt} It
was a crisis about the mass of the electron.  We know like charges
repel.  It is hard to keep electric charge in a small pack because it
repels itself.  On the other hand, we know the electron is basically
point-like.  Our best limit is that the ``size'' of the electron is less
than something like $10^{-17}$~cm.  The problem is that, if you want
to keep the charge in such a small pack, you need a lot of energy.  A
na\"{\i}ve guess is that you need at least
\begin{equation}
  \label{eq:1}
  \Delta E \sim \frac{\alpha}{r_e} \sim 1~{\rm GeV}
  \frac{10^{-17}~{\rm cm}}{r_e}.
\end{equation}
But we know we can't afford it.  The energy carried by an electron is
just $E = mc^2 = 0.511$~MeV, nowhere close to what we need.  In fact,
the best we can do is to pack the charge down to about $10^{-13}$~cm,
which is the so-called classical radius of the electron.  In other
words, the classical theory of electromagnetism breaks down around
this distance scale, and we cannot discuss physics below
$10^{-13}$~cm.  We can't get started!

But we don't talk about this problem anymore, because there was a
resolution.  Antimatter came to the rescue.  We solved the crisis by
doubling the number of particles.  Here is how it works.  

\begin{figure}[t]
  \center
  \includegraphics[scale=0.5]{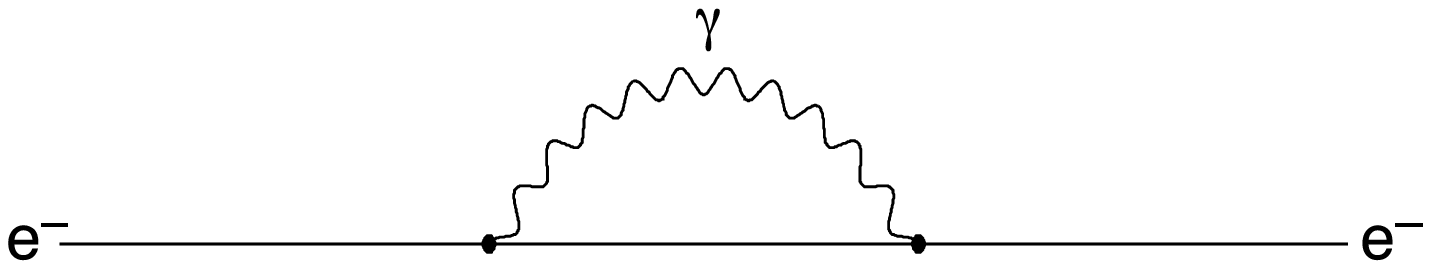}
  \includegraphics[scale=0.5]{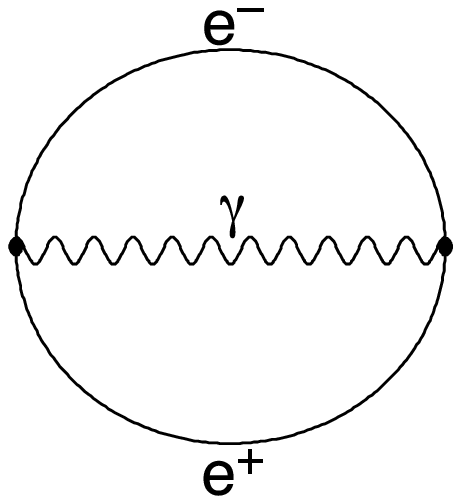}
  \includegraphics[scale=0.5]{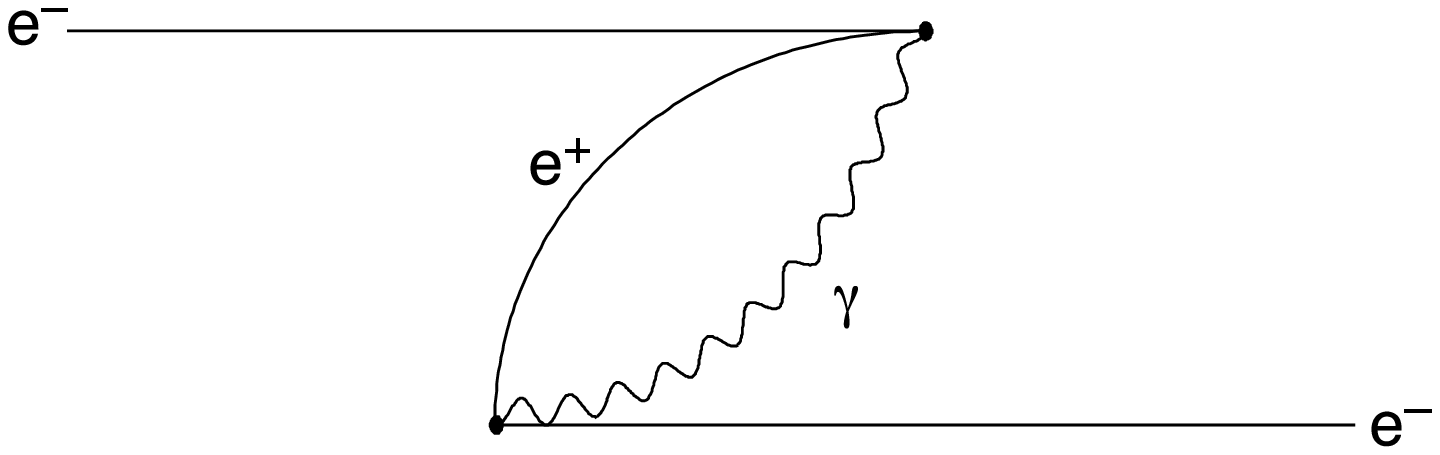}
  \caption{Top: The electron sees the Coulomb field it created itself.  Center:
    The vacuum is full of ``bubbles'' in which an electron-position
    pair is created spontaneously and annihilate back to the vacuum within
    the time allowed by the uncertainty principle.  Bottom: An electron
    may decide to annihilate the positron in the ``bubble'' while the
    electron originally in the ``bubble'' remains as a real
    particle.\label{fig:electron}}
\vspace{-0.4cm}
\end{figure}

The electron creates a Coulomb field around itself, and it feels its
own field.  Namely, it repels itself (Fig.~\ref{fig:electron}, top).
But we discovered antimatter.  Moreover, we discovered that the world
is quantum mechanical.  Once you have these two ingredients, there is
an inevitable consequence.  The ``vacuum'' we see isn't empty at all.
It constantly creates pairs of electrons and positrons, together with
a photon.  Of course, energy conservation forbids it, but quantum
mechanics allows us to borrow energy as long as nobody notices it.
The created pair must annihilate back to the vacuum within the time
allowed by the uncertainty principle (Fig.~\ref{fig:electron},
center).  Such pairs are called ``vacuum bubbles.''

When you place an electron in this fluctuating vacuum, it ``sees'' a
positron nearby.  Sometimes, it decides to annihilate the positron in
the bubble.  Then the electron that was originally a part of the
bubble now remains as a ``real'' particle (Fig.~\ref{fig:electron},
bottom).  It turns out that this process also contributes to the
energy of the electron with a {\it negative}\/ sign, that nearly
exactly cancels the self-repelling energy we were worried about.  The
grand total is roughly
\begin{equation}
  \label{eq:2}
  \Delta m_e c^2 \sim m_e c^2 \times \frac{\alpha}{4\pi} \log
  (m_e r_e).
\end{equation}
This is nice.  First of all, the additional energy you need is
proportional to the original energy (the rest energy $m_e c^2$), and
we are talking about a percentage correction.  Second, even if you
take the smallest size imaginable, namely the Planck size $r_e \sim
10^{-33}$~cm, the size of the correction is only about 10\%.  Now we
can get started to think about physics below $10^{-13}$~cm.

\begin{figure}
  \center
  \includegraphics[width=0.8\columnwidth]{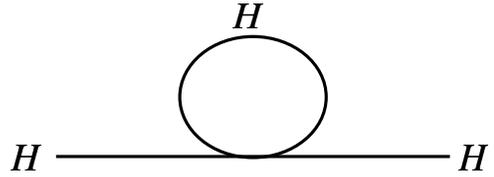}
  \caption{The self-repulsion of the Higgs boson makes it hard to be
    contained in a small size.\label{fig:Higgsself}}
\end{figure}

The problem we are facing now is very similar.  The minute you think
that we are swimming in the Higgs BEC, you should ask if Higgs can be
contained in a small package.  It turns out that the Higgs also repels
itself because of its self-interaction (Fig.~\ref{fig:Higgsself}).  It
requires a lot of energy to contain itself.  The theory breaks down
again, this time around $10^{-17}$~cm.  We are stuck.  We can't get
started to address the big questions.  We can't ``see'' the
interesting physics at shorter distances that answers the big
questions.

\begin{figure}
  \center
  \includegraphics[width=0.8\columnwidth]{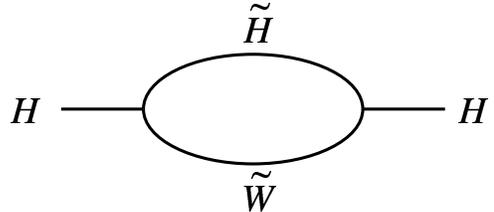}
  \caption{The Supersymmetric attraction diagram cancels the Higgs
    self-repulsion diagram.\label{fig:Higgsself2}}
\end{figure}

One way to solve this problem is to assume that history repeats
itself.  We double the number of particles again.  The new particles
cancel the contribution from the Higgs self-repelling energy
(Fig.~\ref{fig:Higgsself2}).  This is the idea of Supersymmetry, which
makes the Standard Model consistent with whatever physics there is at
shorter distances.  Indeed, the correction to the Higgs energy is
\begin{equation}
  \label{eq:3}
  \Delta m_H^2 \sim \frac{\alpha}{4\pi} m_{SUSY}^2 \log(m_H r_H),
\end{equation}
where $r_H$ is the ``size'' of the Higgs boson.  Of course
Supersymmetry is not the only solution, but it is true that any
solution of this kind appears at the TeV scale.

\begin{figure}
  \center
  \includegraphics[width=0.5\columnwidth]{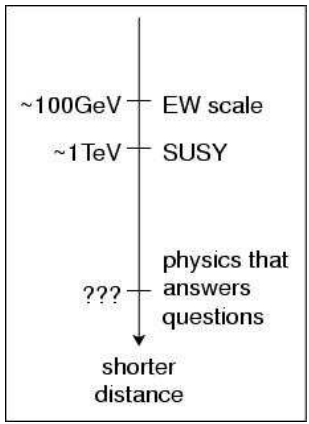}
  \caption{Once the hierarchy problem is solved, we will be allowed to
    talk about physics at shorter distances to address the big
    questions.\label{fig:scales3}}
\end{figure}

Once the hierarchy problem is solved, we can finally get started.  It
opens the door to the answers to the big questions
(Fig.~\ref{fig:scales3}).  The sky clears up and we can start
``seeing'' physics at shorter distances.  An even more interesting
possibility is that the solution itself provides additional probes to
physics at shorter distances.  We will talk about some examples soon.

\begin{figure}
  \center
  \includegraphics[width=0.8\columnwidth]{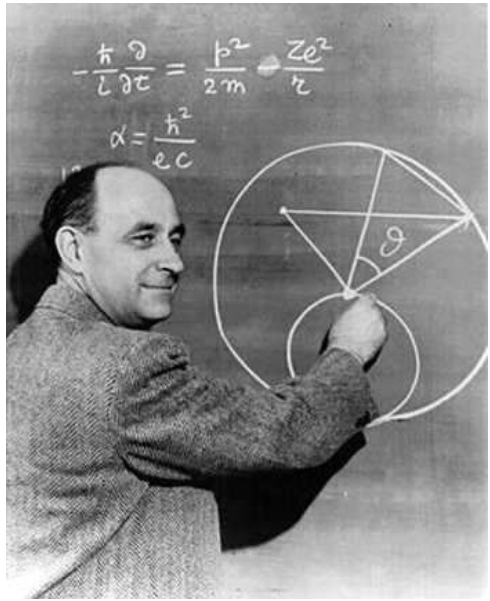}
  \caption{A nice picture of Fermi, except the definition of the fine
    structure constant is wrong.  What was he
    thinking?\label{fig:Fermi}}
\end{figure}

In fact, the importance of the TeV-scale has been known since 1933.
When Fermi (Fig.~\ref{fig:Fermi}) wrote down his theory of nuclear
beta decay, he knew the relevant energy scale: $G_F^{-1/2} \simeq
300$~GeV.  It is truly exciting that we are finally getting to this
energy scale!

For a long time, theorists, including myself, had been talking about
three major directions to solve the hierarchy problem.  One is
Supersymmetry, which I already talked about.  It is the idea that the
history repeats itself.  Just like antimatter solved the crisis
of the electron mass, we double the number of particles.  The second
direction is to learn from Cooper pairs.  The Higgs condensate is a
composite made of two fermions.  This idea is often called
technicolor.  These two ideas are two decades old, and many of you
have witnessed a nearly religious war between the two camps.  The third
direction is relatively recent: physics {\it ends}\/ at a TeV.
The TeV-scale is the ultimate scale of physics where quantum gravity
manifests itself.  It may be superstrings.  We may produce blackholes
at accelerators or by cosmic rays.  This is possible if there are
hidden dimensions curled up in small sizes, somewhere between 10~$\mu$m
to $10^{-17}$~cm.

But the fact that the third direction was proposed relatively recently
suggests that there are many more possibilities we theorists haven't
thought of.  Indeed, just the last two years have seen an outbreak of
new ideas.  The Higgs boson may be like pions in QCD, a
pseudo-Nambu-Goldstone boson.  Models based on this idea are called
the ``little Higgs'' theories\rlap.\,\cite{littlehiggs} Or maybe the Higgs
boson is actually a gauge boson.  Extra-dimensional components of a
gauge field do not appear to have spins to a four-dimensional observer
because they are spinning in extra dimensions.  This idea is sometimes
called Gauge-Higgs Unification\rlap.\,\cite{gauge-higgs} Or maybe there is no
Higgs after all, and the reason why $W$ and $Z$ are massive is because
they are Kaluza--Klein bosons, running along the extra dimension to
acquire their ``rest'' (for a 4D observer) energies.  This idea came
to be known as ``Higgsless'' theories\rlap.\,\cite{higgsless} Most recently, I'm
pushing the idea of ``technicolorful
supersymmetry.''\cite{Murayama:2003ag} You see, I'm pretty ecumenical.

\begin{figure}[t]
\center
\includegraphics[width=\columnwidth]{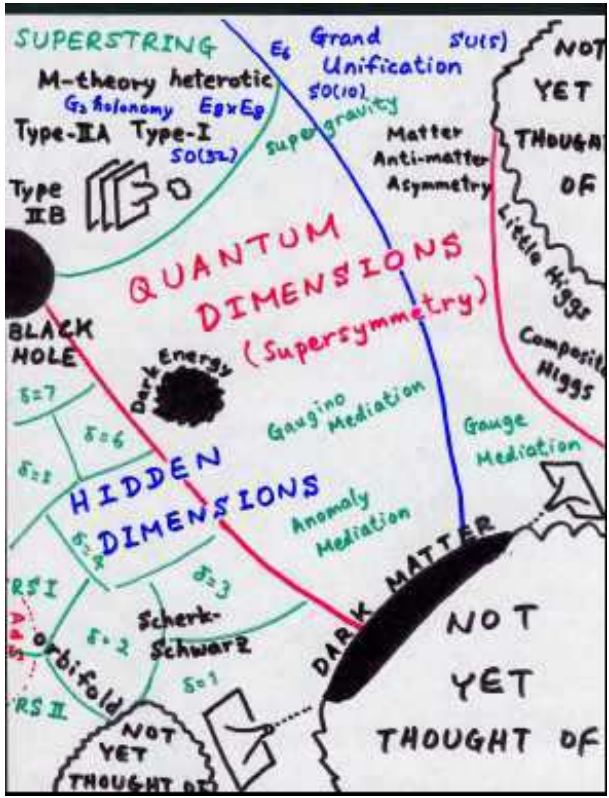}
\caption{The landscape of theories that has many uncharted
  territories.  The task for experiments is to zoom down to a point on
  this map.\label{fig:theory2}}
\vspace{-0.4cm}
\end{figure}

Clearly the landscape of theories is getting more and more complicated
(Fig.~\ref{fig:theory2}).  As the solution to the problem draws near
experimentally, theorists are proposing more possibilities.  It is
increasingly clear that all the theoretical possibilities we have
talked about, have designed the experiments around, and ran Monte Carlo on,
are only a small portion of the land of all theories.  There are many
islands and continents already labeled on the map, but much of the
land is uncharted.

The task for the future experiments is enormous: to zoom in to a point
on the map most of which is still uncharted.  We need to identify the
physics responsible for the Higgs BEC, and it is quite likely that we
haven't thought of the right solution yet.  We are all excited about
the LHC, where we will discover particles and new phenomena that
address this issue.  Many possibilities will be ruled out.  However,
new interpretations will necessarily emerge.  Then the race will be
on.  Theorists come up with new interpretations.  Experimentalists
exclude new interpretations.  It will be a {\it long}\/ period of
elimination.  As is always the case, the crucial information is in
the {\it details}\/.  We would like to elucidate the physics by
reconstructing the Lagrangian of the ``true theory'' term by term from
measurements. 

In this process, the absolute confidence behind our understanding is
crucial, especially when we witness a major discovery.  Just for the
sake of discussion, let us say that Supersymmetry happens to be the
``true theory.''  It is relatively easy to reach, what I'd like to
call, ``New York Times-level confidence.''  We will see a headline
like ``{\it The Other Half of the World Discovered}\/.''  But
everybody in this auditorium knows that there is a long way to go from
this level of confidence to the other level of confidence, which I'd
like to call ``Halliday--Resnik-level confidence.''  It will take an
incredible level of confidence to put a paragraph like this one in the
freshman physics textbook:
\begin{quote}
  \it We have learned that all particles we observe have unique
  partners of different spin and statistics, called superpartners,
  that make our theory of elementary particles valid to small
  distances.
\end{quote}
Upon seeing this slide, one of my colleagues in Berkeley was impressed
by the fact that Halliday and Resnik can turn something as exciting as
the discovery of Supersymmetry into something this dry and dull.
Well, {\it that}\/ wasn't my point.  My point is that we need to go
through many detailed, precise, unambiguous measurements for us to
reach this level of confidence.

\begin{figure}[t]
  \center
  \includegraphics[width=0.5\columnwidth]{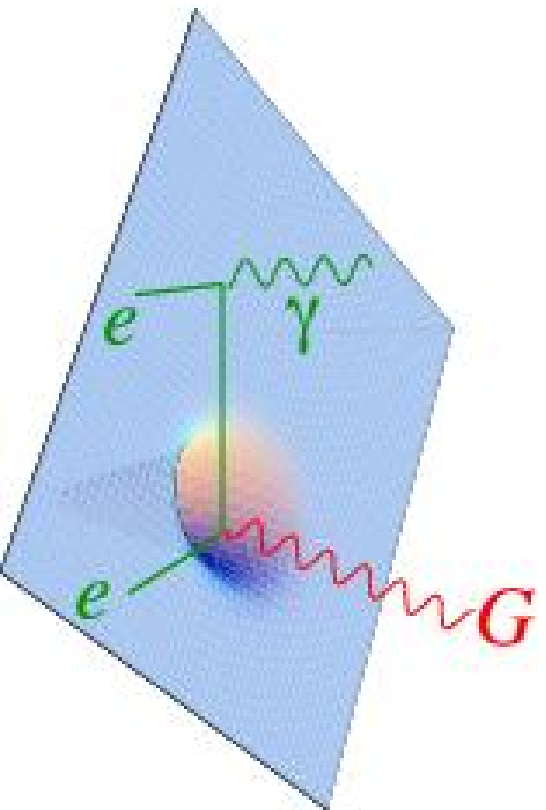}
  \includegraphics[width=0.8\columnwidth]{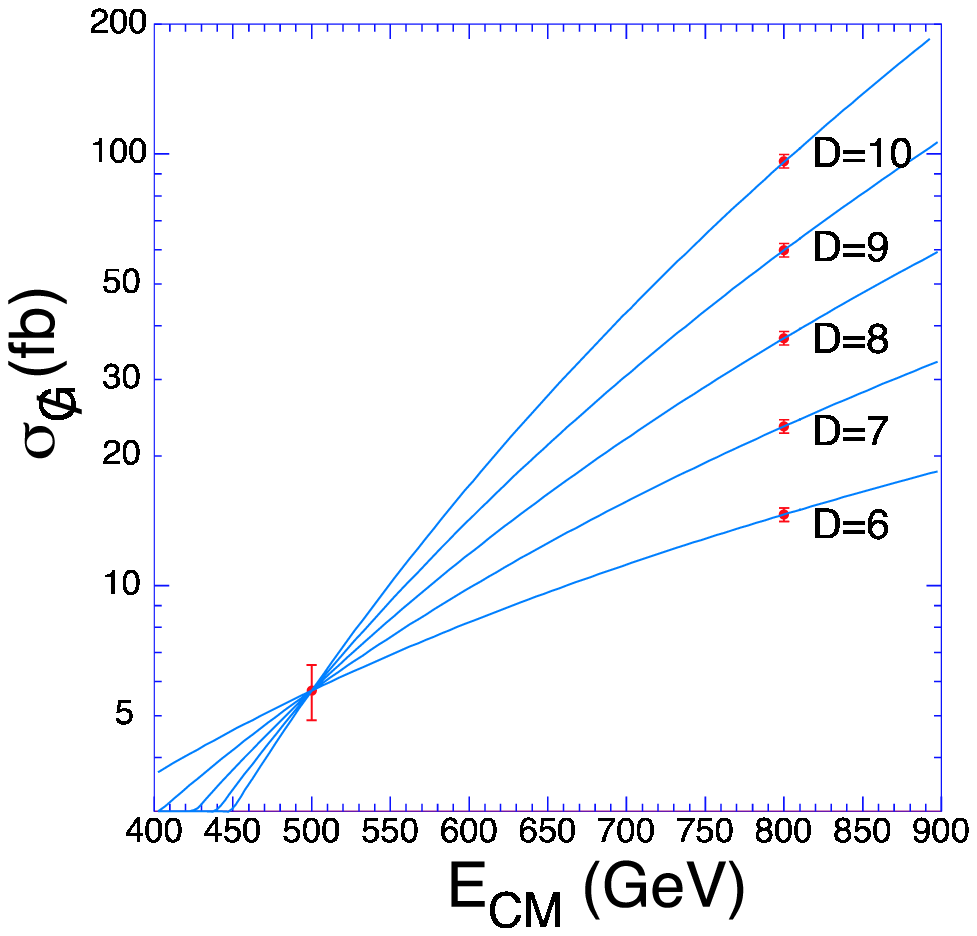}
  \caption[]{Top: Emission of a graviton into the hidden dimensions. Bottom: The
    energy dependence of the rates for various number of
    dimensions\rlap.\,\cite{TESLA}\label{fig:hiddendimensions}}
\end{figure}

Again for the sake of discussion, let us say that hidden dimensions
happen to be the ``true theory.''  We will see events where the
high-energy collisions on our three-dimensional sheet will produce
some particles we can see and other particles that disappear into the
extra dimensions, such as the graviton (Fig.~\ref{fig:hiddendimensions},
top).  Then we find that the energy and momentum are not balanced
apparently.  There is clearly something exciting going on.  However,
such a discovery wouldn't establish the theory.  We'd like to know how
many of such extra dimensions there are, for instance.  One way to
address this question is to measure the rates of this kind of events
at two different energies at an $e^+ e^-$ Linear Collider.  The energy
dependence of the rates can tell us the number of extra dimensions
(Fig.~\ref{fig:hiddendimensions}, bottom).

\begin{figure}[t]
\center
\includegraphics[width=\columnwidth]{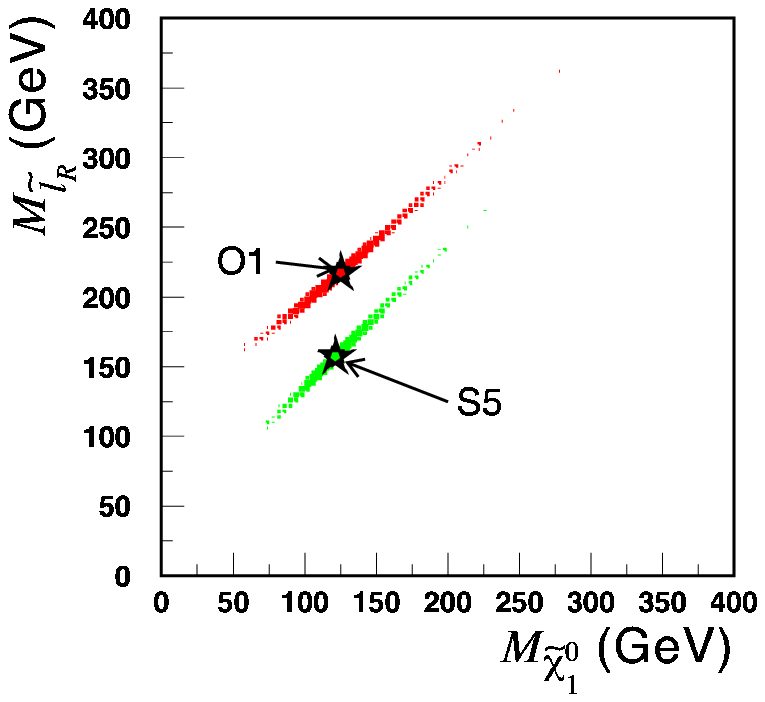}
\caption[]{Precision measurement of superparticle masses at the
  LHC\rlap.\,\cite{Bachacou:1999zb}\label{fig:SUSYmass}}
\includegraphics[width=\columnwidth]{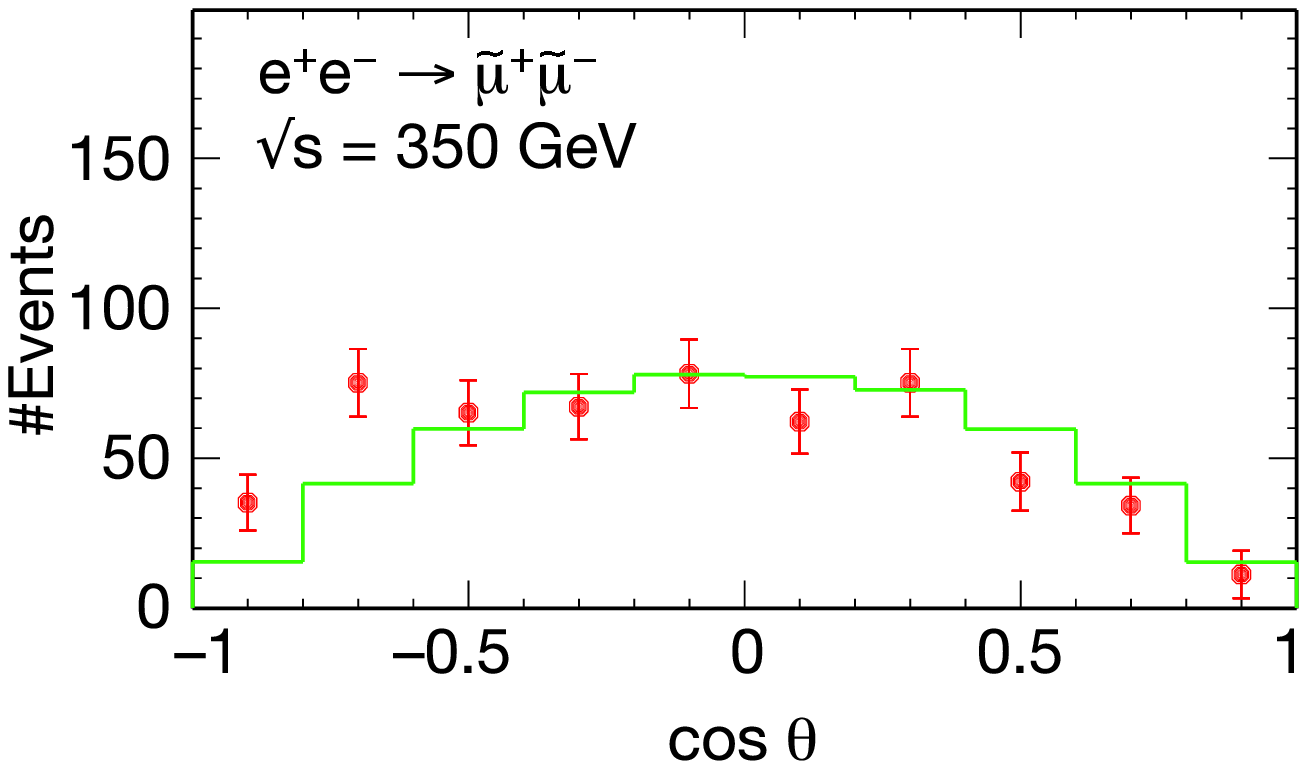}
\caption[]{Test if the smuon has spin zero at the
  LC\rlap.\,\cite{Tsukamoto:1993gt} The spin one case would show an
  upside down angular distribution.\label{fig:smuspin}}
\end{figure}

Let us pick Supersymmetry again.  In this case, the Tevatron and/or LHC
will expand our sensitivity in the parameter space greatly beyond
where we are, and many precise measurements will be performed at the
LHC (Fig.~\ref{fig:SUSYmass}).  However we will still like to know if
the new particles truly have the same quantum numbers as the particle
we already know, with their spins differing by 1/2.  Again the Linear
Collider can determine the quantum numbers and spins, and if they have the
correct couplings, etc. (Fig.~\ref{fig:smuspin}).  This way, we will
establish Supersymmetry with absolute confidence.

\section{Heaven}

I now turn to the questions from the Heaven.  One of the major results
this year reported at this Symposium is the study of cosmic microwave
background anisotropies by the WMAP satellite.  From a global fit, they have reported
precise measurements of important cosmological parameters that
include\cite{Spergel:2003cb}
\begin{eqnarray*}
  &&h = 0.71 \pm 0.04, \\
  &&\Omega_M h^2 = 0.135 \pm 0.009, \\
  &&\Omega_b h^2 = 0.0224 \pm 0.0009, \\
  &&\Omega_{\it total} = 1.02 \pm 0.02.
\end{eqnarray*}
This is yet another big step in precision cosmology.  To me, the
most important information is that the case for non-baryonic dark matter
is now as strong as 12$\sigma$: $|\Omega_M - \Omega_b| h^2 = 0.113 \pm
0.009$.  

People have looked for dim stars or big planets that make up Dark
Matter in the halo of our galaxy, dubbed MACHOs (Massive Compact Halo
Objects).  The search resulted in a strong upper limit on the halo
fraction of such astronomical objects\rlap.\,\cite{MACHO} Instead, we are led
to WIMPs (Weakly Interacting Massive Particles).  They are stable
heavy particles produced in the early Universe when the temperature was as
high as their mass.  As the universe cooled, the temperature was so
low that they were no longer created.  They started to annihilated
with each other, but as the universe expands, they saw fewer and fewer
of each other and beyond some point they could no longer find each
other to annihilate.  In this way, they are left-overs from the near
complete annihilation.  The amount of energy density left over
is\cite{KT}
\begin{equation}
  \Omega_M = \frac{0.756(n+1) x_f^{n+1}}{g^{1/2} \sigma_{\it ann} M_{Pl}^3}
  \frac{3s_0}{8\pi H_0^2} \approx 
  \frac{\alpha^2/({\rm TeV})^2}{\sigma_{\it ann}}.
\end{equation}
It is very interesting that weakly coupled (as weak as $\alpha$)
particles at the TeV scale can provide the correct energy density to
explain the Dark Matter.

\begin{figure}[t]
\center
\includegraphics[width=\columnwidth]{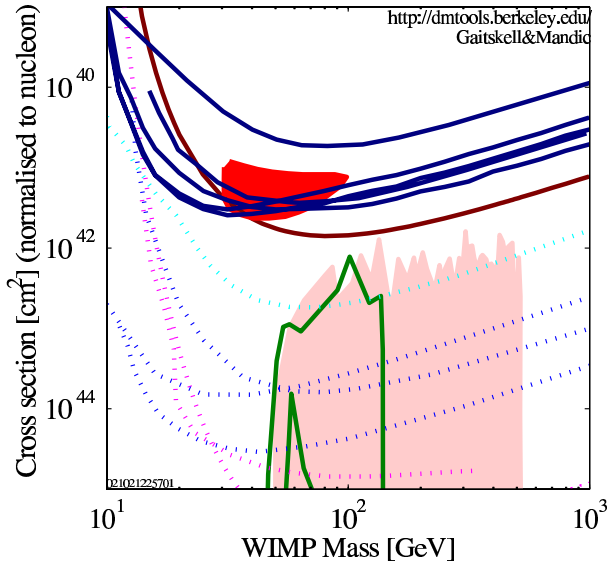}
\caption[]{The next-generation Dark Matter search experiments will get
  into the interesting portion of the WIMP parameter
  space\rlap.\,\cite{dmtools}\label{fig:dmtools}}
\vspace{-0.4cm}
\end{figure}

A stable, weakly-coupled particle would be an excellent candidate for
Dark Matter.  Actually, it should be very weakly coupled because
ordinary neutrinos would be too strongly coupled and are excluded by
the negative search results.  There are no such candidate particles in
the Standard Model.  The candidate most talked about is the Lightest
Supersymmetric Particle (LSP), which is the superpartner of the photon
or $Z$ in most models.  Indeed, the direct search experiments so far
have made only a small foray, but the next generation experiments will
take a significant bite out of the interesting part of the parameter
space (Fig.~\ref{fig:dmtools}).  This way, we will know that Dark
Matter is indeed there floating in the halo of our galaxy.  On the
other hand, we would also like to know what it is.  For this purpose,
we'd like to produce ample quantities of Dark Matter in the laboratory
to study its properties in detail.

\begin{figure}
\center
\includegraphics[width=\columnwidth]{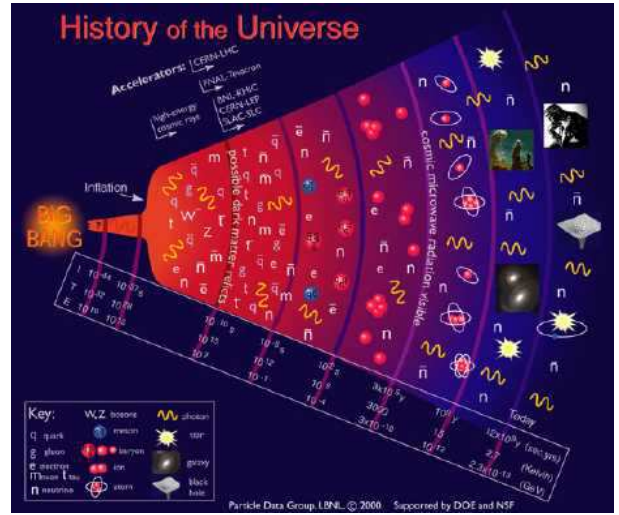}
\caption{Schematic history of Universe.  We have a pretty
  good grip on physics back to about a second after the Big Bang
  thanks to CMB and nucleosynthesis.  The quark-gluon plasma is
  physics back at $10^{-5}$~sec.  An agreement between
  accelerator-based data on Dark Matter and cosmological data would
  provide understanding much closer to The Beginning, back to
  $10^{-12}$~sec after the Big Bang.\label{fig:universe_original}}
\end{figure}

I have argued that the Dark Matter is likely be a TeV-scale
electrically neutral weakly interacting particle.  There are many such
candidates: Lightest Supersymmetric Particle, Lightest Kaluza--Klein
particle in universal extra dimension, etc.  Given that I expect new
particles at the TeV-scale to address the ``Hell'' problems, it is
quite conceivable that one of those particles is stable (or long-lived
enough) to be the Dark Matter. If so, it will be accessible at
accelerators, such as the LHC and LC.  Precision measurements of its mass
and couplings to other particles at LHC and LC will allow us to
calculate its cosmic abundance.  If that calculation based on
accelerator experiments turns out to agree with the cosmological
observations, it would be a major triumph of modern physics.  We will
understand the universe all the way back to when it was only about
$10^{-12}$~sec old after Big Bang (Fig.~\ref{fig:universe_original})!

The Dark Energy is even more mysterious and we should be ready for
more surprises.  One big question is why we seem to see nearly equal
amounts of Dark Energy and Dark Matter now.  This is the notorious
``Why Now?''  problem.  We seem to live at a very special moment in
the evolution of universe.  It almost feels like we are stepping back
from the heliocentric view of Copernicus to the geocentric world of
Ptolemy.  We physicists all hate the idea that we are special.

\begin{figure}
\center
\includegraphics[width=\columnwidth]{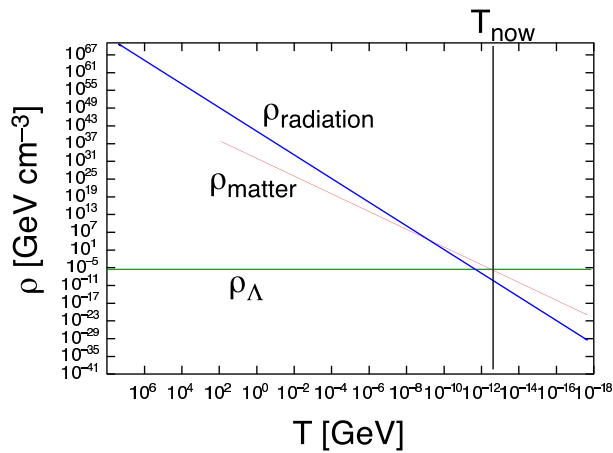}
\caption[]{The triple coincidence of three energy
  densities\rlap.\,\cite{Arkani-Hamed:2000tc}\label{fig:coincidence}} 
\end{figure}

Given that the problem is so big, it is useful to step back a little
bit and look at the situation globally.  Then we find that it is not
just Dark Matter and Dark Energy; the ``radiation,'' which basically
refers to CMB photons and neutrinos, has a similar energy density as
well (Fig.~\ref{fig:coincidence}).  It is actually a triple
coincidence problem.  We have three lines with different slopes that
meet at a single point.  Leaving $O(1)$ numerical constants aside, the
radiation energy density is $\rho_{\rm rad} \sim T^4$, while the Dark
Matter energy density is $\rho_M \sim m^2 T^3/M_{Pl}$, where $m \sim
1$~TeV gives the correct amount as we have seen earlier.  In order for
the Dark Energy to meet with both of them, we need the Dark Energy
density to be $\rho_\Lambda \sim ({\rm TeV}^2/M_{Pl})^4$.  Indeed, the
observation suggests $\rho_\Lambda \approx (2~{\rm meV})^4$, while
${\rm TeV}^2/M_Pl \approx 0.5~{\rm meV}$, tantalizingly close.  It
looks like figuring out TeV-scale physics is crucial for the Dark
Energy problem, too.

\begin{figure}
\center
\includegraphics[width=\columnwidth]{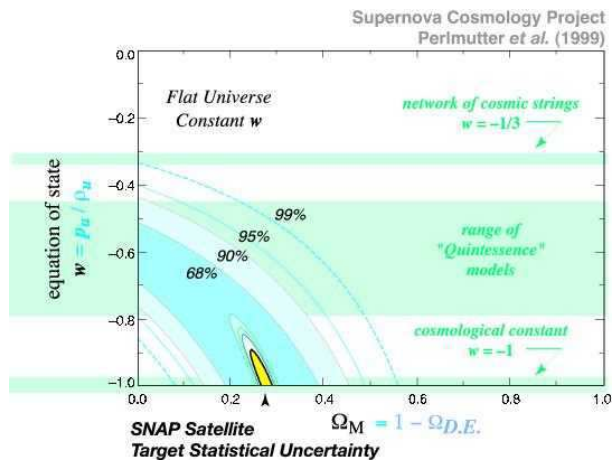}
\caption[]{The projective accuracy of the equation of state of the Dark
  Energy by SNAP\rlap.\,\cite{SNAP}\label{fig:SNAP}}
\end{figure}

The parameter we would most like to measure about the Dark Energy is
its equation of state.  Marc Kamionkowski once told me that the
``equation of state'' is a misnomer.  It is not an equation, but
rather a ratio of the pressure to the energy density $w = p/\rho$.
Due to some reason, it is called the {\it equation}\/ of state, but it
is just a number.  In any case, the cosmological constant corresponds
to $w=-1$, while an evolving dynamical system typically has $w > -1$.
A dedicated high-statistics study of high-redshift supernovae,
complemented by the study of nearby ones to pin down the systematic
issues would be extremely useful: such as SuperNova Acceleration Probe
(SNAP) using a dedicated satellite.  It will determine the ``equation
of state'' at a high accuracy.  My favorite candidate for Dark Energy,
a frustrated network of domain walls\cite{Friedland:2002qs} that leads
to $w=-2/3$, will be cleanly distinguished from the cosmological
constant once SNAP happens (Fig.~\ref{fig:SNAP}).  Once we know the
equation of state, we will get the first glimpse of the nature of the
Dark Energy.  Where to go from there will depend on what we find.

\section{Vertical}

Now on to the Vertical Questions.

Einstein once asked a very simple question.  Is there an underlying
simplicity behind the vast phenomena in Nature?  He dreamed of finding
a {\it unified}\/ description of all phenomena we see.  But he failed
to find a unified theory of electromagnetism and his theory of
gravity, general relativity.

Indeed, trying to come up with a more universal, more fundamental,
more unified theory is in the blood of all of us physicists.  An early
example of unification is Sir Newton: he unified apples and planets.
It was a revolutionary thought: the same law of physics applied to
both terrestrial bodies, like an apple, and celestial bodies, like
planets.  Out of this unification came two important theories,
Newton's law of mechanics, and the inverse-square law of gravity.  A
more familiar example is Maxwell, who unified electric and magnetic
forces.  At the time of Einstein, there were also strange phenomena in
atomic physics that led to quantum mechanics.  In addition, there were
even more mysterious phenomena in nuclear physics, such as
$\alpha$-decay, $\beta$-decay, and $\gamma$-decay.

\begin{figure}
\center
\includegraphics[width=\columnwidth]{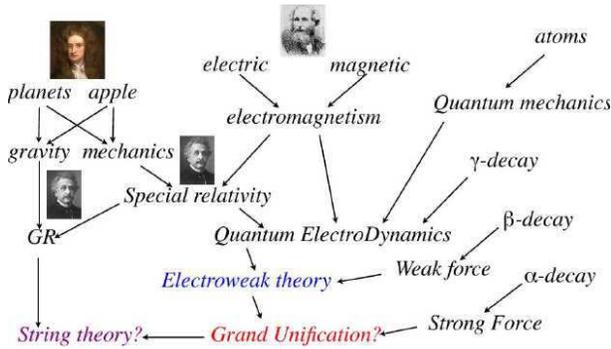}
\caption{A brief history of unification in physics.}
\end{figure}

Later, there was an important unification in physics which somehow
people don't talk about much.  It is Quantum ElectroDynamics
(QED), that unifies special relativity, electromagnetism, quantum
mechanics, and some other phenomena such as nuclear $\gamma$-decay.
It is an incredibly successful theory that predicts the magnetic
moment of the electron down to its twelfth digit.  It is equally
incredible that experiments can measure it down to its twelfth digit,
and all twelve digits agree with each other.  This is a great triumph
for the general idea of unification in physics.

The other phenomena led to discoveries of new forces.  The nuclear
$\beta$-decay was the first manifestation of the weak force, while the
$\alpha$-decay was that of the strong force.  We are now just about to achieve
the next layer of unification, between QED and the weak force.
Beyond that, we are still at the stage of speculation.  The strong
force may be further unified with the electroweak forces into a single
force; it is called the grand unification.  We also would like to see
gravity unified with the other forces.  Currently the best bet is
string theory.

\begin{figure}
\center
\includegraphics[width=\columnwidth]{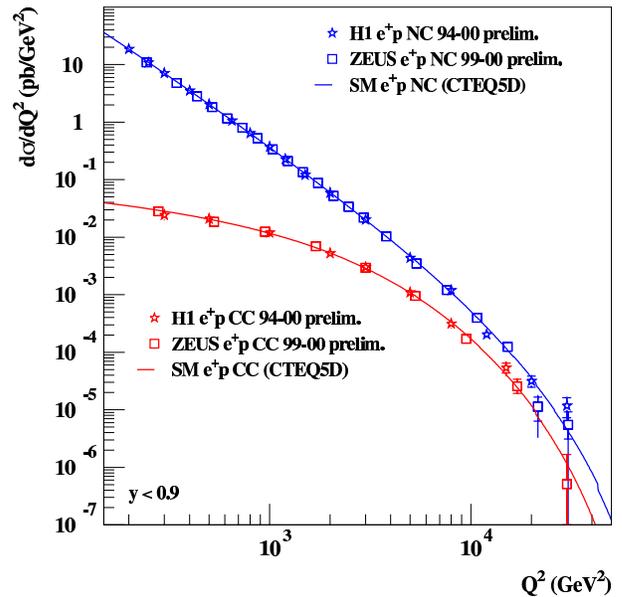}
\caption[]{We are heading towards unification of the electromagnetism
  and the weak force\rlap.\,\cite{HERA} \label{fig:HERA}}
\end{figure}

We are indeed just about to achieve the next layer of unification.
Figure~\ref{fig:HERA} shows the strengths of electromagnetic
and weak forces as a function of the energy scale.  The first
manifestation of the weak force, nuclear $\beta$-decay, was
measured at much lower energies, off the scale in Fig.~\ref{fig:HERA}, where the strengths of
the two forces were many orders of magnitude different.  However our
predecessors figured out that they are supposed to be of the same
kind; an amazing insight.  After many decades we inched up in energy,
and are finally approaching the energy scale where they indeed become
the same.  It has been a long-term goal since the 1960's and we are getting
there!  However the important missing link is the Higgs boson as we
talked about already.

Beyond the unification of electromagnetism and the weak force at the
TeV-scale, how do we gain any information about the next layer, the
grand unification?  We have all seen during the Symposium that the
strengths (gauge couplings) of the three forces, $SU(3)$, $SU(2)$, and
$U(1)$, appear to become equal at a very high-energy scale
$10^{16}$~GeV, if the Standard Model is Supersymmetric.  The energy
scale appears so remote that we may not gain any further information.
However, if Supersymmetry is discovered, and the masses of the new
particles are measured to a high precision by combining the data from
the LHC and the LC, we will have a quantitative test of grand
unification.  The superpartners of the gauge bosons, gauginos, should
have masses that unify at the same energy scale where the gauge
couplings unify.  This is a highly non-trivial test of whether forces
unify.  If this happens, we would definitely want to see proton decay!
This is a wonderful example of how, once the hierarchy problem is
solved, the solution itself will provide new probes to physics that
directly address the big questions.

\begin{figure}[t]
\center
\includegraphics[width=\columnwidth]{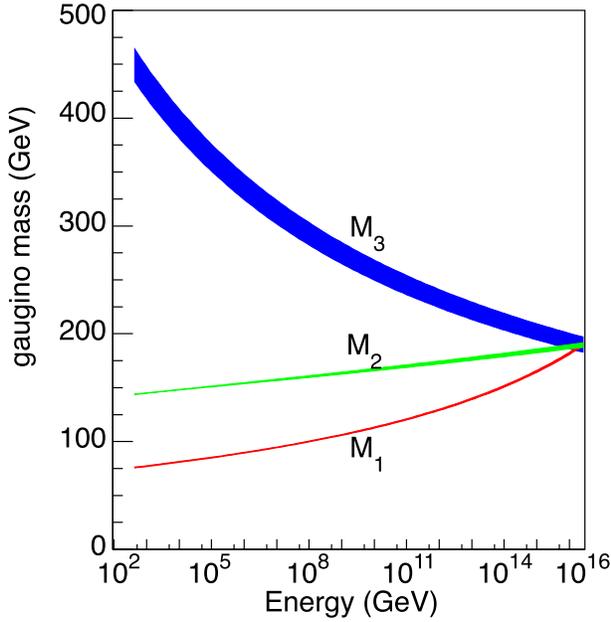}
\caption[]{If Supersymmetry is found, it will provide a further probe to
  shorter distance scale physics, such as testing grand unification
  using the gaugino masses\rlap.\,\cite{Martyn:1999tc}}
\end{figure}

\begin{figure}[t]
\center
\includegraphics[width=\columnwidth]{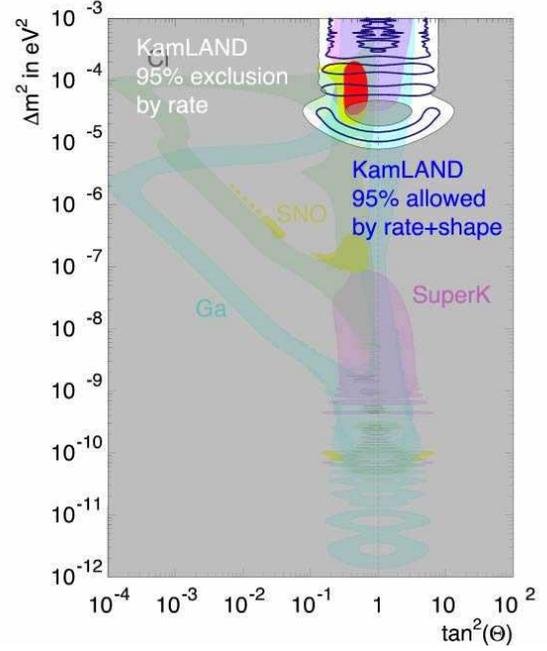}
\caption[]{Solar neutrino data and reactor data converged on the Large
  Mixing Angle solution\rlap.\,\cite{hitoshi}\label{fig:hitoshi}}
\end{figure}

Once the hierarchy problem is solved, we can systematically look for
effects from physics at high energies.  They can be parameterized as
effective operators added to the Standard Model,
\begin{equation}
  {\cal L} = {\cal L}_{\rm SM} + \frac{1}{\Lambda} {\cal L}_5
  + \frac{1}{\Lambda^2} {\cal L}_6 + \cdots
\end{equation}
where $\Lambda$ is the high energy scale of new physics.  The effects
in ${\cal L}_5$ are suppressed by a single power of the high energy
scale, ${\cal L}_6$ by two powers, etc..  What terms there can be have
been classified systematically by Weinberg, and there are many terms
suppressed by two powers:
\begin{eqnarray}
  {\cal L}_6 &\supset& QQQL,\, \bar{L} \sigma^{\mu\nu} W_{\mu\nu} H e,\,
  W_\nu^\mu W_\lambda^\nu B_\mu^\lambda,\, \nonumber \\
  & & \bar{s}d \bar{s}d,\, (H^\dagger D_\mu H) (H^\dagger D^\mu H), \cdots\, .
\end{eqnarray}
The examples here contribute to proton decay, $g-2$, the anomalous triple
gauge boson vertex, $K^0$--$\overline{K}^0$ mixing, and the
$\rho$-parameter, respectively.  It is interesting that there is only
one operator suppressed by a single power:
\begin{equation}
  {\cal L}_5 = (LH) (LH).
\end{equation}
After substituting the expectation value of the Higgs, the Lagrangian
becomes
\begin{equation}
  {\cal L} = \frac{1}{\Lambda} (LH)(LH)
  \rightarrow \frac{1}{\Lambda} (L\langle H\rangle)(L\langle H\rangle)
  = m_\nu \nu \nu,
\end{equation}
nothing but the neutrino mass.

Therefore the neutrino mass plays a very unique role.  It is the
lowest-order effect of physics at short distances.  This is a very
tiny effect.  Any kinematical effects of the neutrino mass are
suppressed by $(m_\nu / E_\nu)^2$, and for $m_\nu \sim 1$~eV which we
now know is already too large and $E_\nu \sim 1$~GeV for typical
accelerator-based neutrino experiments, it is as small as $(m_\nu /
E_\nu)^2 \sim 10^{-18}$.  At first sight, there is no hope to
probe such a small number.  However, any physicist knows that
interferometry is a sensitive method to probe extremely tiny effects.
For interferometry to work, we need a coherent source.  Fortunately
there are many coherent sources of neutrinos in Nature: the Sun,
cosmic rays, reactors (not quite Nature), etc..  We also need
interference for an interferometer to work.  Fortunately, there are
large mixing angles that make the interference possible.  We also need
long baselines to enhance the tiny effects. Again fortunately there
are many long baselines available, such as the size of the Sun, the
size of the Earth, etc..  Nature was very kind to provide all the necessary
conditions for interferometry to us!  Neutrino interferometry, a.k.a.
neutrino oscillation, is a unique tool to study physics at very high
energy scales.  Indeed, the na\"{\i}ve interpretation of the neutrino
oscillation results we heard about during this conference suggests
$\Lambda \sim 10^{15}$~GeV!  This gives an important look at the physics of grand
unification.

\section{Horizontal}

The Horizontal Questions are about the flavor.  As we witnessed during
this conference, this is a historic era in flavor physics.  In the
lepton sector, Cowan and Reines detected neutrinos from a nuclear
power reactor back in 1956, but we hadn't learned much about the
nature of neutrinos for decades.  In 1998, SuperKamiokande announced
the discovery of oscillation in atmospheric neutrinos\rlap.\,\cite{atmos} In
2002, SNO established the flavor conversion in solar
neutrinos\rlap.\,\cite{solar} Later the same year, KamLAND decided the
solution to the solar neutrino problem
(Fig.~\ref{fig:hitoshi})\rlap.\,\cite{reactor}

\begin{figure}
\center
\includegraphics[width=\columnwidth]{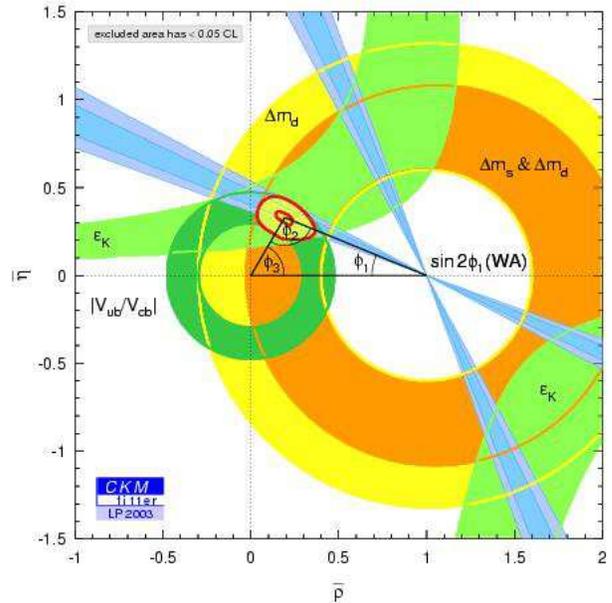}
\caption[]{Consistency of various CKM parameter measurements shows the
  success of Kobayashi--Maskawa
  theory\rlap.\,\cite{CKM}\label{fig:CKMfit}}
\end{figure}

In the quark sector, the progress is similarly spectacular.  Back in
1964, Fitch and Cronin discovered indirect CP-violation in
$K^0$--$\overline{K}^0$ mixing.  Again we didn't learn much beyond it
for decades.  Then in 1998, CPLEAR established $T$-violation in
the same system\rlap.\,\cite{Angelopoulos:1998dv} In 1999, KTeV and NA48
established the direct CP-violation,
$\varepsilon'/\varepsilon$\rlap.\,\cite{epsilonprime} In 2001, BaBar and
Belle established indirect CP-violation in the $B_d$ system, the first CP-violation
in a system other than kaons\rlap.\,\cite{sin2beta} These results
combined to confirm the Kobayashi--Maskawa theory of CP-violation
(Fig.~\ref{fig:CKMfit}).

But there are even more questions to be answered.  The main question
is this: {\it What distinguishes different generations?}\/ Three
generation of particles have the same quantum numbers, but they look
very different.  They have masses hierarchically different by many
orders of magnitude.  They mix rather little.  Both hierarchical
masses and small mixings go against our ``common sense'' in quantum
mechanics.  If two states share exactly the same set of quantum
numbers, you expect them to have similar energy levels, and you expect
them to mix a lot.  The lack of both suggests that there is an ordered
structure behind the flavor.

Many theorists including myself think that there are probably hidden
flavor quantum numbers that distinguish different generations.  A
quantum number means a new symmetry according to Noether's theorem: a
flavor symmetry.  This new symmetry must allow the top quark Yukawa
coupling because it is $O(1)$.  On the other hand, it forbids all the
other Yukawa couplings so that all other quarks start out massless.
But this symmetry must be only approximate, and is broken a little.
This small symmetry breaking allows other Yukawa couplings, generating
small and hierarchical Yukawas.  Because different generations have
different quantum numbers, they are not allowed to mix.  Again the
small symmetry breaking allows them to mix by small amounts.

Here is a toy model of a simple $U(1)$ flavor
symmetry\rlap.\,\cite{Haba:2000be} I basically introduce a new charge to all
particles.  This symmetry is broken by a small parameter $\langle
\epsilon \rangle \sim 0.02$ of charge $-1$.  Let me assign charges to
quarks and leptons in the following way,
\begin{eqnarray}
  \label{eq:4}
  {\bf 10} (Q, u_R, e_R) (+2, +1, 0)\\
  {\bf 5}^* (L, d_R) (+1, +1, +1).
\end{eqnarray}
Here, I used grand-unified terminology of decuplet and quintet, and
the three charges refer to the first, second, and third generation,
respectively.  This charge assignment keeps the top quarks, both left-
and right-handed, neutral, and the top quark Yukawa coupling is
allowed, while all other entries of the Yukawa matrices are
forbidden.  However, using $\epsilon$, we can fill in other entries as
well.  We find
\begin{eqnarray}
  \label{eq:5}
  M_u \sim \left( 
    \begin{array}{ccc}
      \epsilon^4 & \epsilon^3 &\epsilon^2 \\
      \epsilon^3 & \epsilon^2 & \epsilon \\
      \epsilon^2 & \epsilon & 1
    \end{array} \right), \\
  M_d \sim \left( 
    \begin{array}{ccc}
      \epsilon^3 & \epsilon^3 &\epsilon^3 \\
      \epsilon^2 & \epsilon^2 & \epsilon^2 \\
      \epsilon & \epsilon & \epsilon
    \end{array} \right), \\
  M_l \sim \left( 
    \begin{array}{ccc}
      \epsilon^3 & \epsilon^2 &\epsilon \\
      \epsilon^3 & \epsilon^2 & \epsilon \\
      \epsilon^3 & \epsilon^2 & \epsilon
    \end{array} \right).
\end{eqnarray}
It is easy to find the hierarchical mass eigenvalues,
\begin{eqnarray}
  \label{eq:6}
  & & m_u : m_c : m_t \sim m_d^2 : m_s^2 : m_b^2  \nonumber \\
  & & \sim
  m_e^2 : m_\mu^2 : m_\tau^2 \sim \epsilon^4 : \epsilon^2 : 1,
\end{eqnarray}
which works pretty well especially given how simple the toy model is.
The mixing angles are also suppressed by powers in $\epsilon$, and
they all work out within a factor of 5 or so.

It is exciting that new data from neutrinos are already providing
significant new information about flavor symmetries.  As you know,
neutrino data has been full of surprises.  All mixing angles are
large, except for $U_{e3}$ which has not been measured.  In
particular, the atmospheric neutrino mixing appears maximal.  Two
mass-squared splittings are not very different, $\Delta m^2_{\it
  solar}/\Delta m^2_{\it atmospheric} \sim 1/30$.  The hierarchy in
masses rather than (masses)$^2$ is the square root of this,
$\sqrt{1/30} = 0.2$.  This isn't much of a hierarchy.  Now we can ask
the question of whether there is a symmetry or structure behind the
neutrinos.

As far as we can tell, we don't need any symmetry or structure behind
neutrinos, unlike the quark and charged leptons.  If you run a Monte
Carlo of random complex three-by-three matrices with the seesaw
mechanism, you find that the maximal mixing is the most likely outcome
if plotted against $\sin^2 2\theta$, and the peak in $\Delta m^2_{\it
  solar}/\Delta m^2_{\it atmospheric}$ is about $1/10$.  Apparently no
particular structure in the neutrino mass matrix is needed.  I called
this observation ``anarchy'' in neutrinos.  Actually, the charge
assignments I discussed earlier did not distinguish the three
generation of neutrinos at all; and we do expect anarchy that is
consistent with the current data.

\begin{figure}[t]
  \center
  \includegraphics[width=\columnwidth]{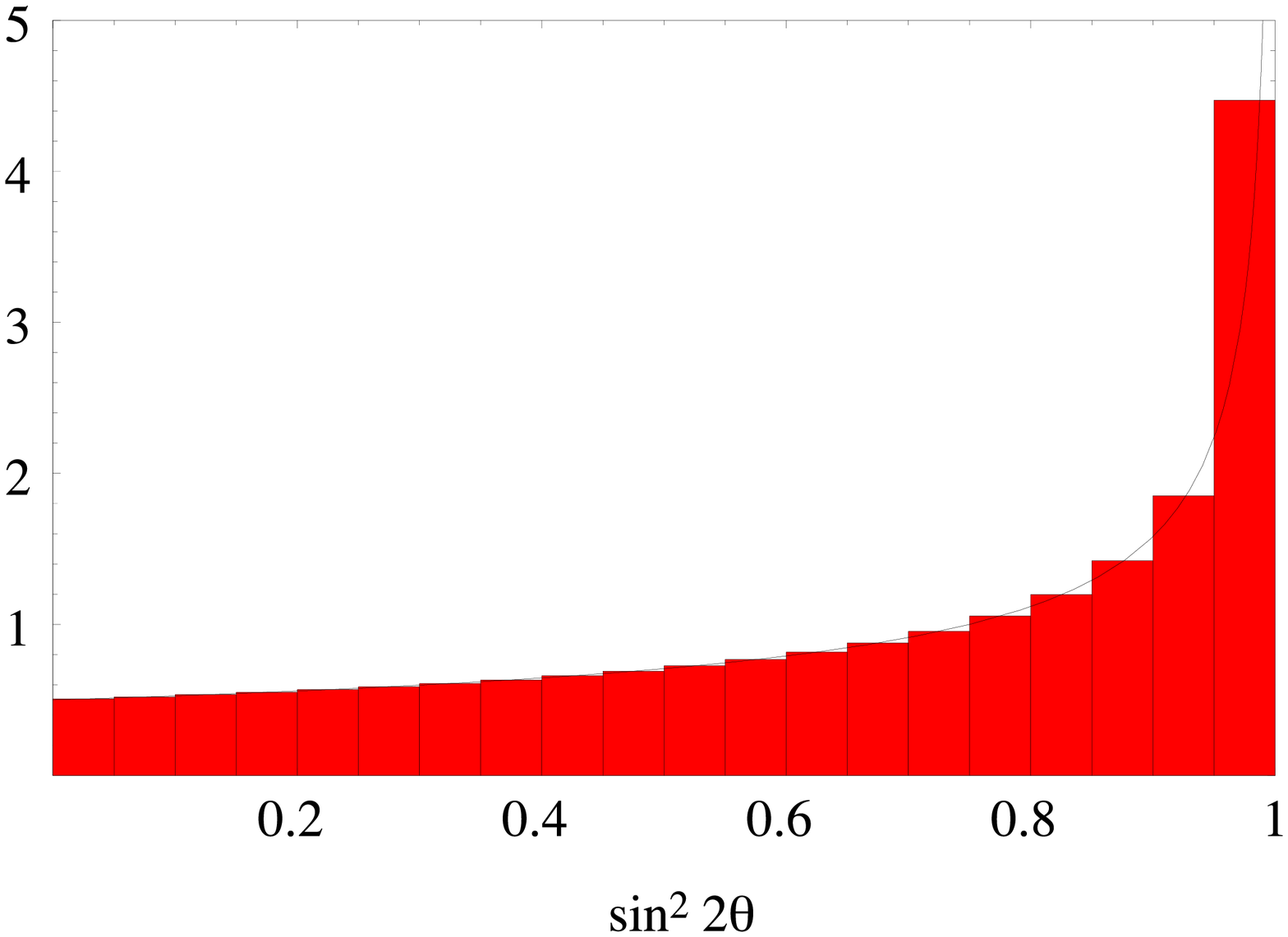}
  \includegraphics[width=\columnwidth]{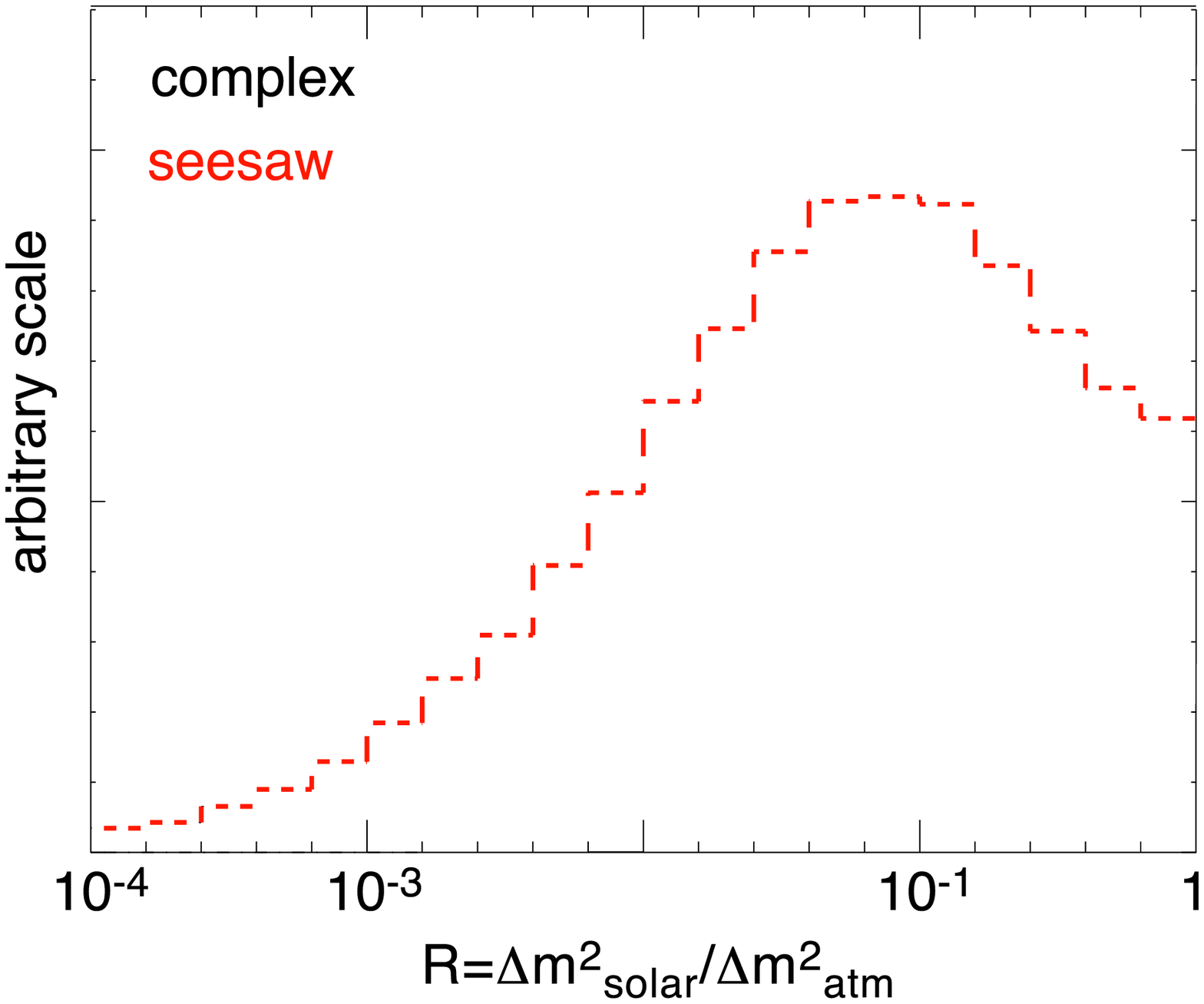}
  \caption[]{Random three-by-three matrices show distributions quite
    consistent with the observed patterns of neutrino masses and
    mixings\rlap.\,\cite{Haba:2000be} Top: $\sin^2 2\theta_{23}$.
    Bottom: $\Delta m^2_{\it solar}/\Delta m^2_{\it
      atmospheric}$.\label{fig:anarchy}}
\end{figure}

Of course there are other proposals to understand the neutrino masses
and mixings together with quarks and charged leptons.
Table~\ref{tab:AFM} shows a list of proposed models of flavor
symmetries as of October 2002.  By December, KamLAND excluded the
third and fourth flavor symmetries because they predicted other
solutions to the solar neutrino problem.  The survivors will be put to
further tests soon.  They predict different orders of magnitude for
$\theta_{13}$, $O(1)$, $O(\lambda)$, or $O(\lambda^2)$.  If a more
precise measurement of $\theta_{13}$ turns out to give $\sin^2
2\theta_{23} = 1.00 \pm 0.01$, we would probably want a reason why it
is {\it so}\/ maximal, implying a new symmetry in the neutrino sector.

\begin{table}
\center
\includegraphics[width=\columnwidth]{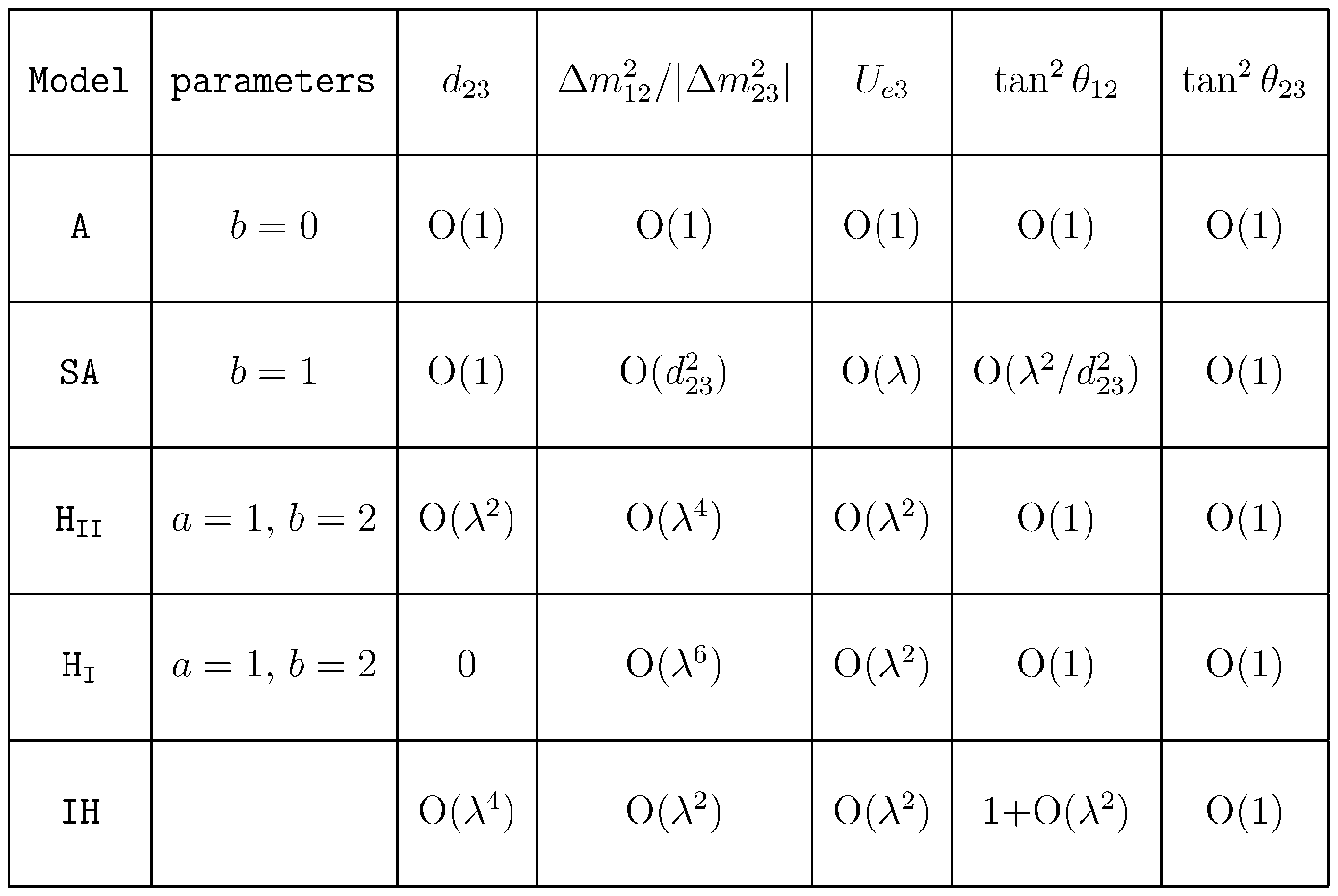}
\caption[]{Compilation of different flavor symmetry models and their
  predictions for neutrino masses and mixings as of Oct.
  2002\rlap.\,\cite{Altarelli:2002sg}  The third and fourth rows were
  excluded by Dec. 2002.  The next benchmark is $U_{e3}$.\label{tab:AFM}}
\end{table}

\begin{figure}[t]
\center
\begin{displaymath}
  \left( 
    \begin{array}{c}
      \tilde{s}_R \\ \tilde{s}_R \\ \tilde{s}_R \\
      \tilde{\nu}_\mu \\ \tilde{\mu}
    \end{array} \right)
  \leftrightarrow
  \left( 
    \begin{array}{c}
      \tilde{b}_R \\ \tilde{b}_R \\ \tilde{b}_R \\
      \tilde{\nu}_\tau \\ \tilde{\tau}
    \end{array} \right)
\end{displaymath}
\caption{The large $\nu_\mu \rightarrow \nu_\tau$ mixing suggests a
  large mixing of the whole $SU(5)$ multiplets and also of their
  superpartners.\label{fig:mixing}}
\end{figure}

We'd like to push this program further to narrow down the choice of
flavor symmetries.  We basically need more and more flavor parameters.
In fact, any TeV-scale physics would have a new flavor structure that
affects flavor physics significantly.  Let me take Supersymmetry as an
example.  Squarks and sleptons come with their own mass matrices, in
addition to quark and lepton mass matrices.  Off-diagonal elements in
squark/slepton mass matrices violate flavor.  Therefore, a flavor
symmetry that distinguishes different generations will automatically
suppress the off-diagonal elements.  If we can probe such small flavor
violations in Supersymmetric loop diagrams, we would like to identify
patterns in them, and eventually deduce the required symmetry behind
them.  Basically, we try to repeat what Gell-Mann and Okubo did in
baryon and meson masses to identify the symmetry behind masses and
mixings.

Different models differ in flavor quantum number assignments.
Different quantum number assignments lead to different consequences
for $\theta_{13}$, the matter effect, CP-violation, $B$-physics,
$K$-physics, Lepton Flavor Violation, proton decay, and practically
anything we can imagine that involves flavor.  This way, we hope to
identify the underlying flavor symmetry.  I admit this is a long shot.
We even don't know the energy scale of the physics of flavor.  It may
turn out to be too remote to access directly in experiments.  But I'd
like to argue that this is not necessarily bad.  In archaeology, you
don't reproduce the events in the laboratory.  But once you have
enough circumstantial evidence of fossils, artifacts, geological
records, etc., that are consistent with a reasonable hypothesis, you
eventually believe it.  It may not be a formal proof at the level
particle physicists are accustomed to, but it is nonetheless the next
best thing.  A good example is the cosmic microwave background.  It is
a wonderfully colorful, sexy fossil, and we can extract so much
information out of it.  We don't recreate the Big Bang, but we have
already learned so much and we will learn even more from the CMB.

I'd like to emphasize that this program will be a collaboration of
energy-frontier experiments and low-energy flavor experiments.  We
need to know the TeV-scale physics so that we know what runs inside the
loops.  We need to know their masses.  Then the flavor data will allow
us to extract flavor violations among the particles in the loops.

\begin{figure}[t]
\center
\includegraphics[width=0.85\columnwidth]{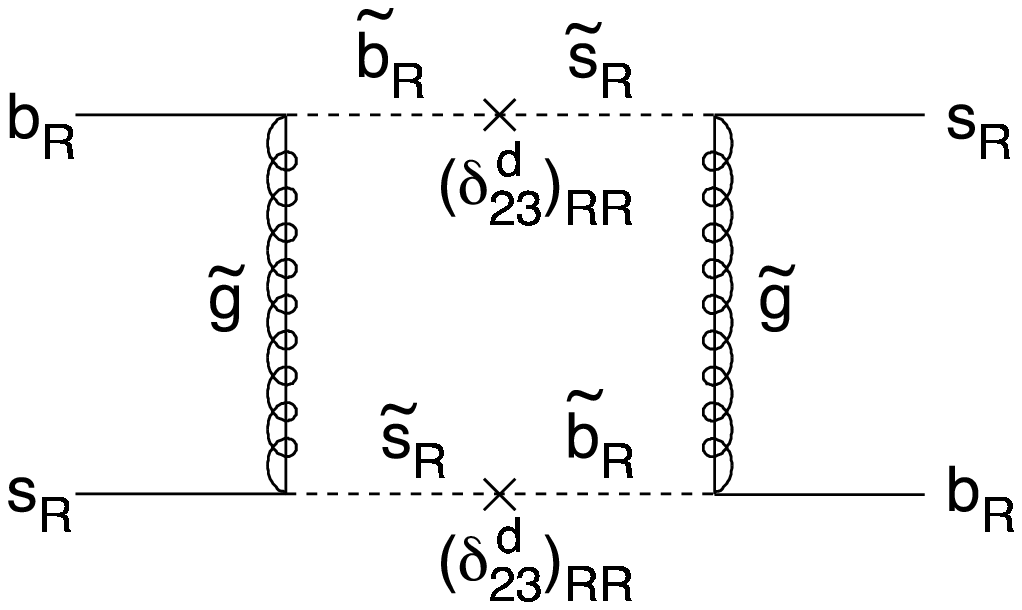}
\includegraphics[width=0.85\columnwidth]{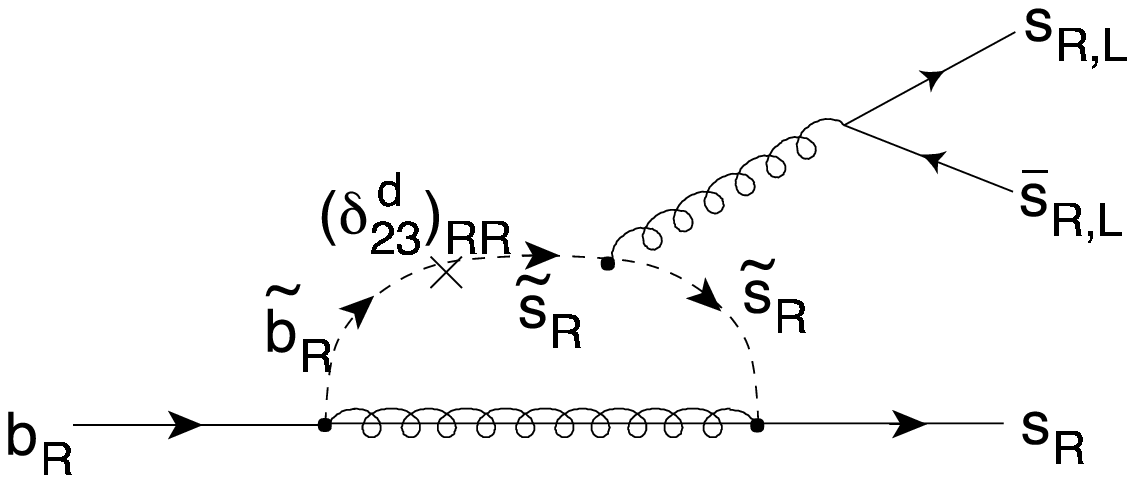}
\includegraphics[width=0.85\columnwidth]{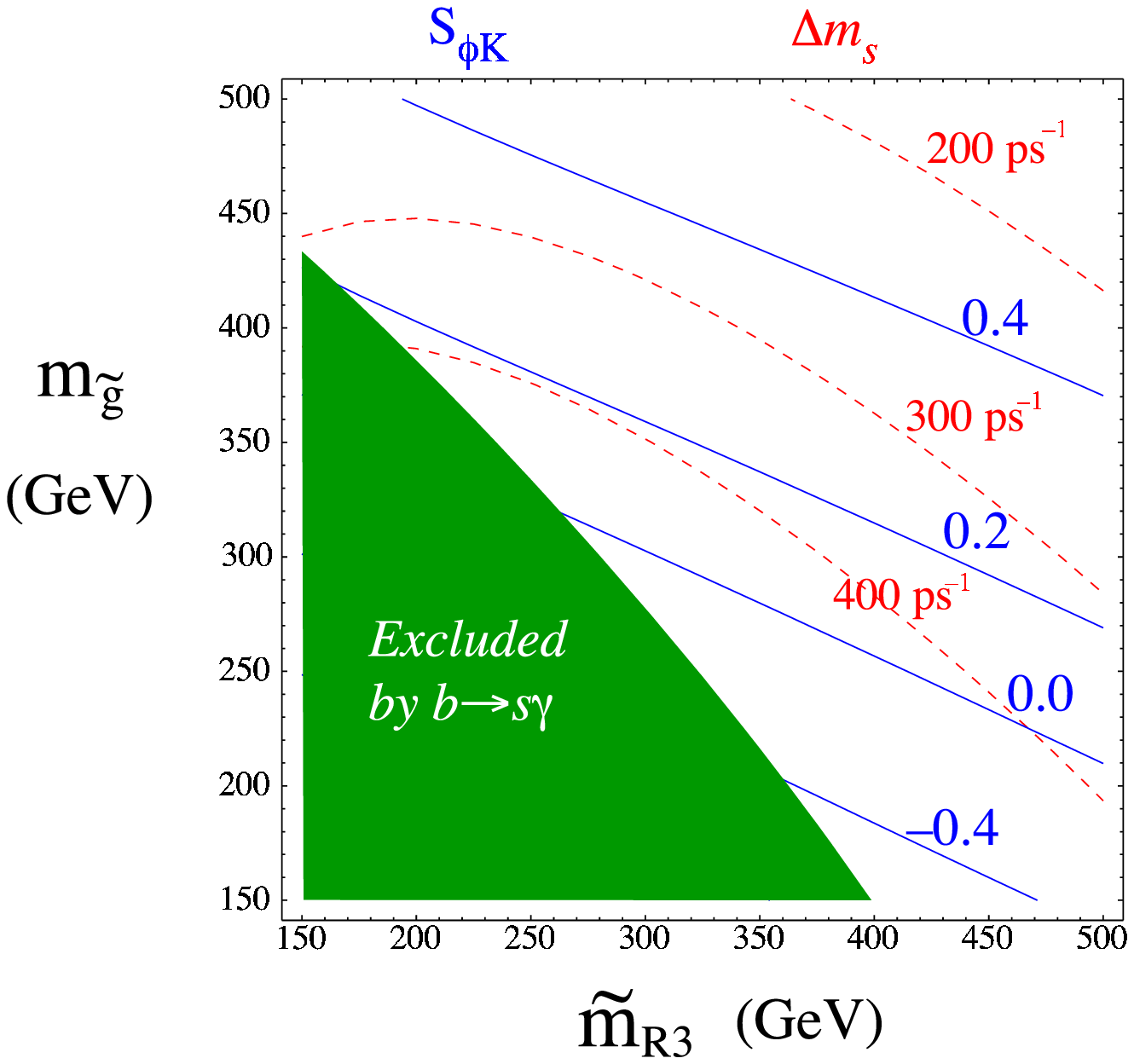}
\caption[]{The impact of large $\tilde{s}_R$--$\tilde{b}_R$ mixing on
  $B$-physics.  Top: possible contribution to the $B_s$ mixing.
  Center: possible contribution to the $B_d \rightarrow \phi K_S$
  decay.  Bottom: $S_{\phi K}$ in solid lines, $\Delta m_s$ in dotted
  lines, and the constraint from $b\rightarrow s\gamma$ in shaded
  region\rlap.\,\cite{Harnik:2002vs}\label{fig:B}}
\end{figure}
\begin{figure}
  \center 
  \includegraphics[scale=0.5]{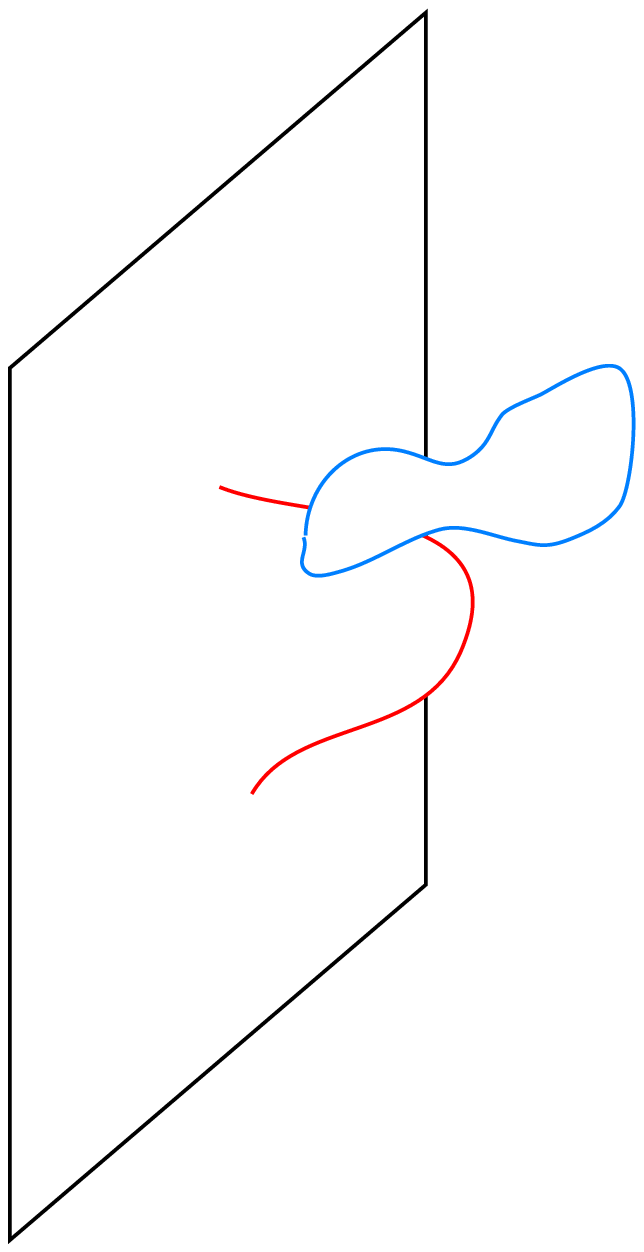}
  \includegraphics[scale=0.5]{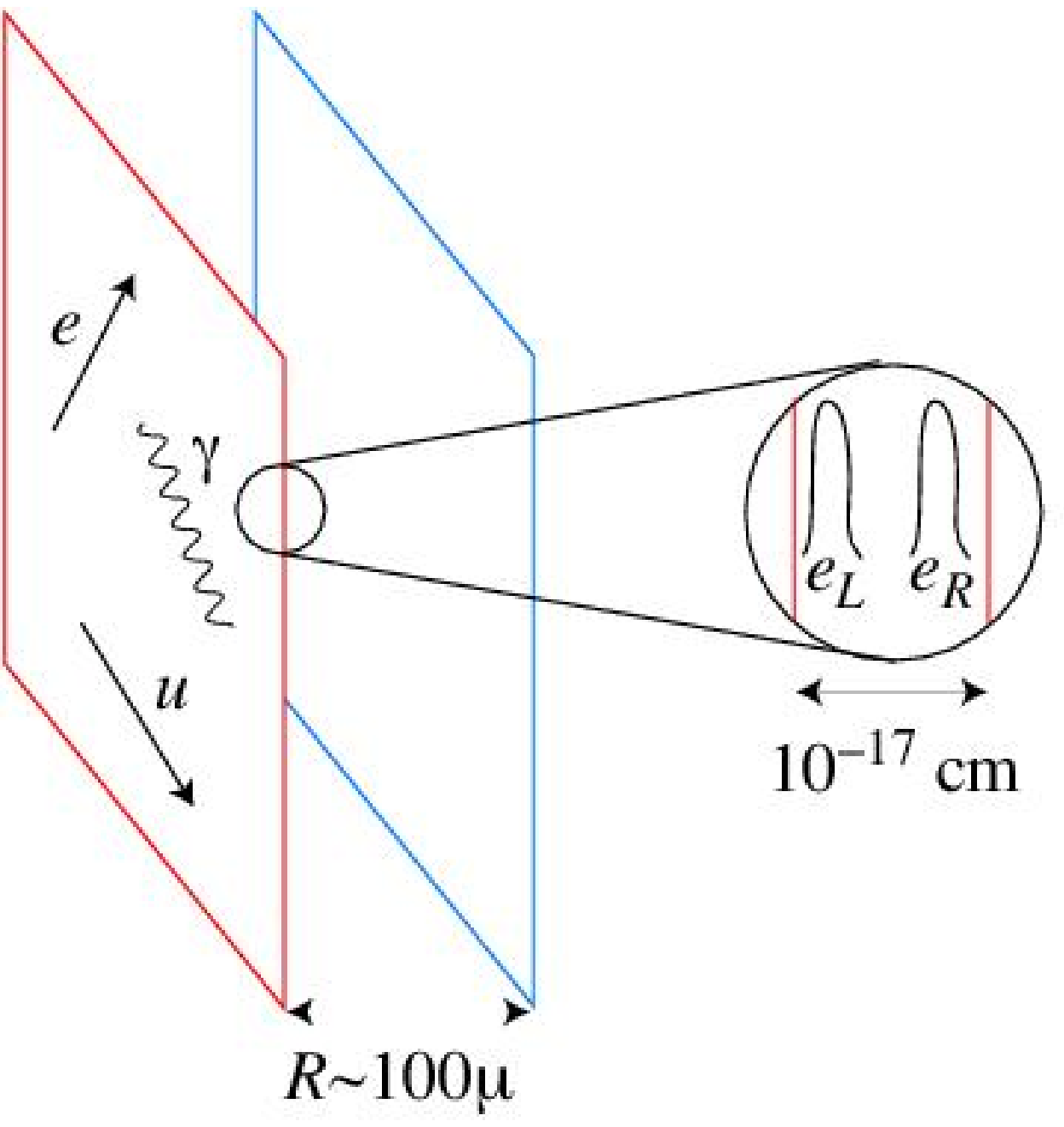}
  \includegraphics{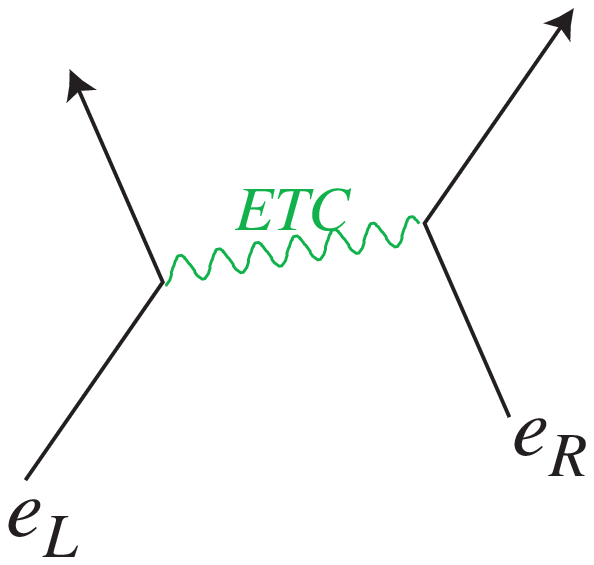}
  \caption{Different views on the origin of flavor symmetry depending on
    the outcome of the TeV-scale physics.  Top: string origin in
    Supersymmetric models.  Center: physical dislocation of different
    generations within a fat brane in models with hidden dimensions.
    Bottom: exchange of new massive gauge bosons at 100~TeV scale in
    technicolor models.\label{fig:flavor}}
\end{figure}

Here is one specific example we should pursue\rlap.\,\cite{Chang:2002mq} We'd
like to know if quarks and leptons have a common origin of flavor.  As
already mentioned, one striking discovery was that the $\nu_\mu$ and
$\nu_\tau$ mix {\it a lot}\/, maybe even maximally.  Suppose you make
it grand-unified.  $s_R$ lives in the same multiplet as $\nu_\mu$, and
$b_R$ with $\nu_\tau$.  You'd expect a large mixing between $s_R$ and
$b_R$, too (Fig.~\ref{fig:mixing}).  But mixing among right-handed
quarks completely drops out from the CKM phenomenology because there
is no right-handed charged current (as far as we know).  It looks like
we can't probe this question.  On the other hand, if there is
Supersymmetry, a large mixing between $\tilde{s}_R$ and $\tilde{b}_R$
is physical, and can induce $O(1)$ effects in $b\rightarrow s$
transitions through loop diagrams (Fig.~\ref{fig:B}, top and center).
Especially in leptogenesis that relies on CP-violation in the neutrino
sector, we expect CP-violation in $\tilde{s}_R$--$\tilde{b}_R$ mixing
that may show up in $B$-physics.

For example, we may see CP-violation in $B_s$ mixing that can be
studied in $B_s \rightarrow J/\psi + \phi$.  The rates in $B_d
\rightarrow X_s \ell^+ \ell^-$ may differ from the Standard Model and
CP-violation may be seen.  CP-violation in $B_d \rightarrow \phi+K_S$
may be different from that in $J/\psi + K_S$ within all the other
constraints, such as $b \rightarrow s\gamma$ (Fig.~\ref{fig:B},
bottom).  The current situation for $\phi K_S$ is somewhat confusing,
with BaBar and Belle not consistent with each other\rlap.\,\cite{CKM} If they
will settle in the middle, however, that may be the first indication
of the common origin of flavor between quarks and leptons!  I'm very
much looking forward to more data.

After going through this program, suppose we identify a reasonable
flavor symmetry that explains all data.  Then we will be greedy enough
to want to know what physics is behind the flavor symmetry.  In the
case of Gell-Mann--Okubo, once the $SU(3)$ flavor symmetry was
identified, the next step was QCD.  Clearly, we have to be very very
lucky to get to that level.  Nonetheless, it is useful to remember
that the next level will crucially depend on what we find at the
TeV-scale.  For example, if the TeV-scale turns out to be
Supersymmetry, the flavor symmetry may be a consequence of the
anomalous $U(1)$ gauge group with the Green--Schwarz mechanism from string
theory (Fig.~\ref{fig:flavor}, top)\rlap.\,\cite{anomalousU(1)} If it is
hidden dimensions, it may be that the three-dimensional brane we live
on is fat enough so that different generations are physically
dislocated within the brane, providing an effective flavor symmetry
(Fig.~\ref{fig:flavor}, center)\rlap.\,\cite{Arkani-Hamed:1999dc} If it is
technicolor, it may be due to a new broken gauge interaction at the
100~TeV scale (Fig.~\ref{fig:flavor}, bottom).\cite{Eichten:1979ah} I
certainly can't see far enough to know how things will play out.

\section{Conclusion}

What I'm looking forward to seeing in the next twenty years is a synergy
of many different approaches in particle physics.  The big questions
I've listed at the beginning of my talk are all very ambitious
questions.  They are elusive.  There is no guarantee that we can
answer them.

But we know what the main obstacle is.  It is the cloud of the
TeV-scale that is preventing us from obtaining clear views.  And we
are getting there.  We have to make sure that there are many different
approaches diverse enough to determine what is going on at the
TeV-scale.  They will converge to reveal the big picture.  Even though
what I'm saying is ambitious, it is {\it conceivable}\/.  And this
idea of synergy applies to any scenario of TeV-scale physics, as far as
I can tell.

Given all this discussion, the outlook for the next twenty years is:\nobreak
\begin{quote}
  \it Bright!
\end{quote}
\begin{figure}[h]
\center
\includegraphics[width=0.90\columnwidth]{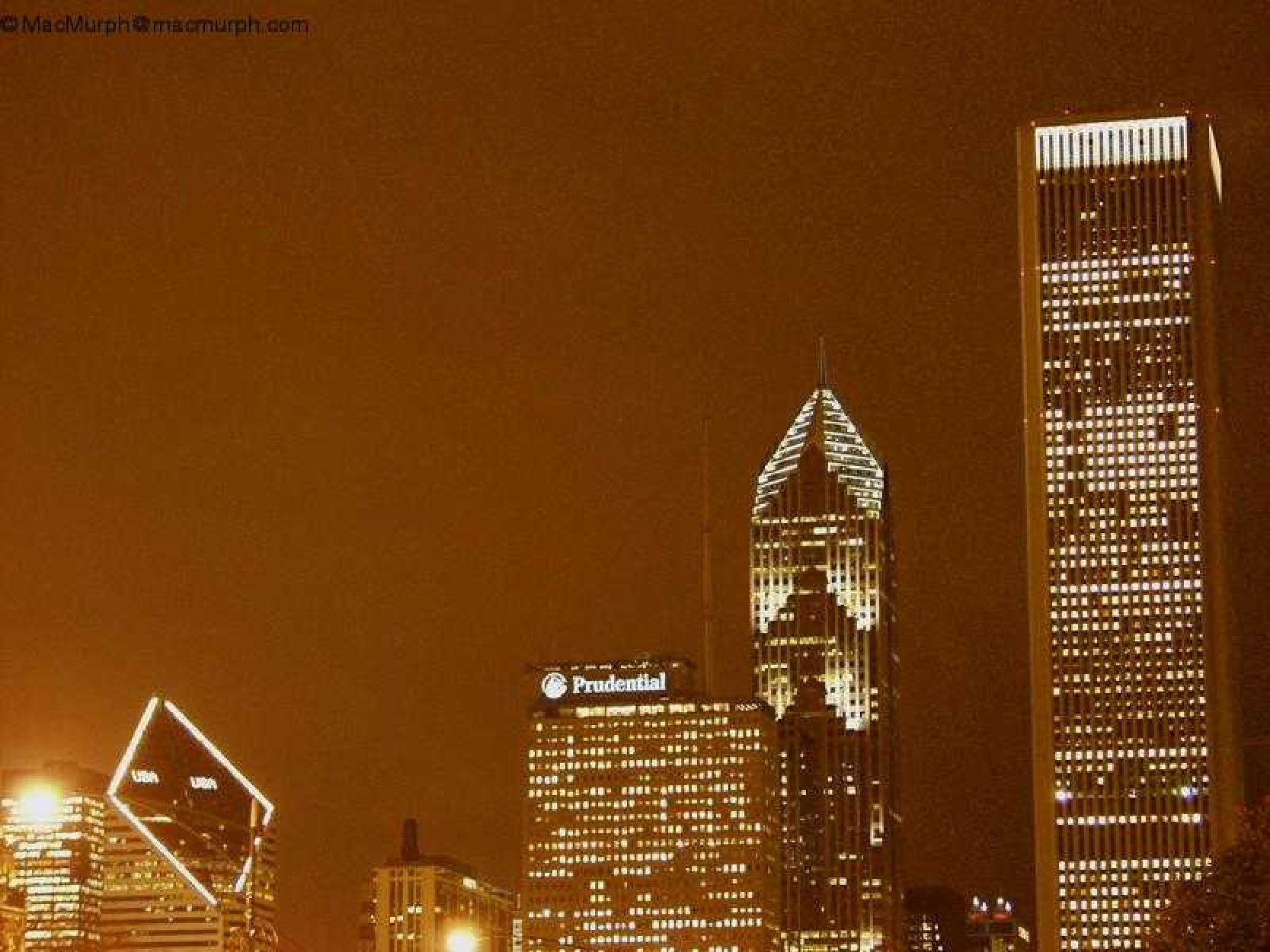}
\end{figure}

\balance

\clearpage
\twocolumn[
\section*{DISCUSSION}
]

\begin{description}

\item[Bennie Ward] (Baylor University \& University of Tennessee):
In your discussion of the hierarchy problem you did not mention the
anthropic principle. Could you please comment?

\item[Hitoshi Murayama{\rm :}]
As Ed said in the previous talk, I don't see the anthropic principle as
the solution to a physical question.  I suppose you can't exclude it,
however.

\item[John Collins] (Penn State):
You said that the Standard Model breaks down at a scale of around a
TeV.  How do you reconcile this with the fact the renormalized Standard Model
is consistent to much higher energies?

\item[Hitoshi Murayama{\rm :}]
It is a matter of definition what you mean by ``breaks down.''  The
Standard Model is certainly consistent as a renormalizable field
theory, and can be applied to arbitrary high energies in that sense.
However, we view it as a low-energy effective field theory rather than
the ultimate theory of everything, and therefore it has an ultraviolet
cut-off.  My definition of the Standard Model breakdown is the fact that
the perturbative corrections exceed the bare Higgs mass-squared
parameter as the cut-off is raised beyond TeV.  It is the same sense as
when Landau and Lifshitz discussed the breakdown of classical
electrodynamics at the classical radius of the electron.

\end{description}

\end{document}